\documentclass[a4paper,11pt]{article}
\pdfoutput=1
\usepackage{jheppub,caption,subcaption}
\DeclareMathOperator\arctanh{arctanh}

\title{A modulated shear to entropy ratio}

\author[a]{O. Ovdat}
\author[a]{and A. Yarom}

\affiliation[a]{Department of Physics, Technion, Haifa 32000, Israel}

\emailAdd{omrieovdat@gmail.com}
\emailAdd{ayarom@physics.technion.ac.il}

\abstract{
We study correlation functions in an equilibrated spatially modulated phase of Einstein-Maxwell two-derivative gravity. We find that the ratio of the appropriate low frequency limit of the stress-stress two point function to the entropy density is modulated. The conductivity, the stress-current and current-stress correlation functions are also modulated. At temperatures close to the phase transition we obtain analytic expressions for some of the correlation functions.}

\begin{document}

\maketitle

\section{Introduction and summary}
The ratio of the shear viscosity to  entropy density in thermally equilibrated systems which are homogeneous, isotropic and described by a holographic dual is given by the celebrated relation \cite{Policastro:2001yc}
\begin{equation}
\label{E:etas}
	\frac{\eta}{s} = \frac{1}{4\pi}\,.
\end{equation}
The relation \eqref{E:etas} has been observed to hold for a wide range of theories \cite{Policastro:2002se,Policastro:2002tn,Herzog:2002fn,Herzog:2003ke,Buchel:2004hw,Parnachev:2005hh,Benincasa:2005iv,Benincasa:2006ei,Adams:2008wt,Herzog:2008wg,Maldacena:2008wh,Mas:2006dy,Son:2006em,Saremi:2006ep,Mateos:2006yd,Ge:2008ak,Landsteiner:2007bd} with a general proof of its validity  given in \cite{Buchel:2003tz,Kovtun:2004de,Buchel:2004qq}. 
The proof of \eqref{E:etas} relies on the system being in a thermally equilibrated isotropic  and homogeneous configuration and on it being described by a dual two-derivative gravity action. If the latter conditions are not met \eqref{E:etas} may fail. For instance, once the 't Hooft coupling of the gauge theory or the rank of its gauge group are finite, the ratio \eqref{E:etas} is corrected \cite{Buchel:2004di,Benincasa:2005qc,Benincasa:2006fu,Brigante:2007nu,Kats:2007mq,Buchel:2008vz,Dutta:2008gf,Brustein:2008xx,Brustein:2008cg,Myers:2009ij,Cremonini:2009sy,Cremonini:2011ej,Cremonini:2012ny}. More recently, \eqref{E:etas} has been shown to break down in non isotropic configurations which are described by two derivative theories of gravity \cite{Erdmenger:2010xm,Basu:2011tt,Erdmenger:2011tj,Rebhan:2011vd,Jain:2014vka,Critelli:2014kra}.

The breakdown of \eqref{E:etas} in  \cite{Erdmenger:2010xm,Basu:2011tt,Erdmenger:2011tj,Rebhan:2011vd,Jain:2014vka,Critelli:2014kra} is not too surprising. Heuristically, the robustness of \eqref{E:etas} can be argued for by observing that the shear viscosity tensor $\eta$ is susceptible only to tensor mode fluctuations of the dual bulk metric which, in an isotropic background, decouple from the other metric fluctuations thereby leading to universal behavior. Once the background is non isotropic there is no reason to believe that \eqref{E:etas} will hold.

This work aims at studying the relation \eqref{E:etas} in a thermally equilibrated, spatially modulated phase of a gauge theory with a $U(1)^3$ anomaly. 
In the pioneering work of \cite{Nakamura:2009tf,Ooguri:2010kt} it was shown that isotropic finite temperature configurations of gauge theories with a gravitational dual with a (large) $U(1)^3$ anomaly become unstable at low temperatures. This instability leads to a modulated phase where translation invariance in one of the spatial directions is broken. (The study of a three space-time dimensional analogue of this instability can be found in, for example, \cite{Donos:2011bh,Bergman:2011rf}.) 

The bulk geometry of the spatially modulated phase discussed in \cite{Nakamura:2009tf,Ooguri:2010kt} was constructed in \cite{Donos:2012wi}.
The key observation of \cite{Donos:2012wi} which allowed for a bulk description of the spatially modulated phase is that the latter possess an $E(2)\times \mathbb{Z}_2$ symmetry ($E(2)$ is the two dimensional Euclidian group) which is a subgroup of the $E(3)$ symmetry of the three dimensional space in which the theory lives in. This $E(2)\times\mathbb{Z}_2$ symmetry of the boundary theory configuration lifts to a Bianchi $VII_0$ symmetry of its asymptotically AdS${}_5$ dual. The fact that the bulk space-time possesses a Bianchi $VII_0$ symmetry significantly reduces the degree of complexity of the problem. The Einstein equations become ordinary differential equations instead of partial differential equations and are therefore easier to manage. An extended classification of the possible symmetries possessed by black branes in asymptotically AdS geometries has been carried out in \cite{Iizuka:2012pn}.

In what follows we will use the setup of \cite{Donos:2012wi} to construct explicit solutions to the Einstein equations in a certain probe limit first discussed in \cite{Ooguri:2010kt}. As we will show, in the probe limit the equations of motion become amenable and allow us to obtain explicit expressions for various correlation functions. In order to present our results we introduce some notation. The bulk action we use is given by
\begin{equation}
\label{E:bulkaction}
	S= \frac{1}{16\pi G_5} \int \left[ \left(R+\frac{12}{L^2}-\frac{1}{4}L^2 F_{MN}F^{MN}\right) + \frac{\gamma}{3} L^3 \epsilon^{ABMNP}F_{AB}F_{MN}A_{P} \right] \sqrt{-g} d^5x\,,
\end{equation}
where $R$ is the Ricci Scalar, $L$ is the AdS radius which we will set to one from now on and $F=dA$ is an Abelian field strength. We have omitted boundary terms which make the variational principle well defined. Capital Roman indices run from $0,\ldots,4$ where the first 4 directions are associated with the coordinate system on the boundary theory. We will refer to the $5$th coordinate as the radial coordinate. The appearance of a Chern-Simons term in the bulk action \eqref{E:bulkaction} implies that the $U(1)$ current dual to the gauge field $A_M$ is anomalous \cite{Witten:1998qj}. Supersymmetry usually implies that $\gamma = \frac{1}{4\sqrt{3}}$ but we will keep it arbitrary in our model. 

As argued for  in \cite{Nakamura:2009tf,Ooguri:2010kt,Donos:2012wi} once $\gamma$ is large enough then there exists a critical temperature $T_c(\gamma)$ for which a non trivial solution to the Einstein equations with Bianchi $VII_0$ symmetry will exist and is more stable than the standard isotropic black brane solution. In the limit where $\gamma$ is very large and $A_{M}/\gamma$ is finite, the matter action decouples from the Einstein-Hilbert action and we obtain a set of equations which can be solved perturbatively in a power series in $1/\gamma$. We will refer to such a scheme for solving the equations of motion as the probe limit.

Within the probe limit we can compute the response of the metric to perturbations. The standard AdS/CFT dictionary relates this response to correlation functions in the dual field theory. We find that the ratio of the imaginary part of the retarded stress-stress correlation functions $G_R^{xy,xy}$ and $G_R^{yz,yz}$ to the entropy density $s$, in the small frequency limit, is given by
\begin{subequations}
\label{E:finalresult}
\begin{align}
\label{E:etasfinal1}
	\lim_{\omega \to 0} \frac{\hbox{Im}(G_R^{xy,xy})/\omega}{s} &= \frac{1}{4\pi} + \frac{t_0}{2\pi \gamma^2} + \mathcal{O}(\gamma^{-3}) \\
\label{E:etasfinal2}
	\lim_{\omega \to 0} \frac{\hbox{Im}(G_R^{yz,yz})/\omega}{s} &= \frac{1}{4\pi} + \frac{t_1+t_2 \cos(4 k x)}{2\pi \gamma^2} + \mathcal{O}(\gamma^{-3})\,,
\end{align}
with $t_i$ dimensionless functions. 
The $x$ coordinate in \eqref{E:etasfinal1} and \eqref{E:etasfinal2} signifies the direction in which translation invariance is broken: if $x,\,y$ and $z$ denote the spatial directions then we use 
\[
	\xi = \partial_x - k (y\partial_z - z \partial_y)
\]
in addition to $\partial_y$ and $\partial_z$ as the generators of the $E(2)$ symmetry. The $\mathbb{Z}_2$ symmetry which we mentioned earlier is associated with a residual rotation symmetry in the $(x,z)$ plane by an angle $\pi$: $x\to-x$ and $z\to-z$. The wavenumber associated with the helical phase is given by $k$ which is determined from thermodynamic considerations. 

We can write expressions for the $t_i$ in  \eqref{E:etasfinal1} and \eqref{E:etasfinal2} in integral form. Close to the phase transition and with a compactified $x$ direction, we can compute most of the $t_i$ analytically. We present such a computation in section \ref{S:Perturbative} after we have introduced some of our notation and methods. A' priori, $t_0$ can be positive or negative (or vanish in certain cases). It would be interesting to test whether, for finite values of $\gamma$, \eqref{E:etasfinal1} can be used as a unitarity bound on the strength of the anomaly. 

In an isotropic configuration the shear viscosity, $\eta$, is determined via a zero frequency limit of a stress-stress correlation function,
\[
	G_R^{\mu\nu,\rho\sigma}(\omega,\vec{q};-\omega',-\vec{q}') = -i \int \theta(t) [T^{\mu\nu}(t,\vec{x}),T^{\rho\sigma}(t',\vec{x}')] e^{i \omega t}e^{-i \vec{q} \cdot\vec{x}} e^{i \omega' t'}e^{-i \vec{q}' \cdot\vec{x}'} dt' dt d^3x d^3x'  \,.
\]
Spatial and time translation invariance imply that
\[
	G_R^{\mu\nu,\rho\sigma}(\omega,\vec{q};-\omega',-\vec{q}')=G_R^{\mu\nu,\rho\sigma}(\omega,\vec{q})\delta(\omega-\omega')\delta(\vec{q}-\vec{q}')\,.
\]
As a result, the zero momentum limit, $\vec{q} \to 0$, of $G_R^{\mu\nu,\rho\sigma}(\omega,\vec{q};-\omega',-\vec{q}')$ is encapsulated in $G_R^{\mu\nu,\rho\sigma}(\omega,0)$. The shear viscosity is determined by the Kubo formula
\[
	\eta = \lim_{\omega \to 0} \frac{\hbox{Im}(G_R^{xy,xy}(\omega,0))}{\omega}\,.
\]
In the helical phase translation invariance is not a symmetry so the zero momentum limit of the stress-stress correlator is given by $G_R^{\mu\nu,\rho\sigma}(\omega,\vec{q};-\omega,0)$ which characterizes the response of the energy momentum tensor to a time dependent metric perturbation. The correlators in expressions \eqref{E:etasfinal1} and \eqref{E:etasfinal2} are evaluated in the latter zero momentum limit. We give a detailed account of this point in section \ref{S:Correlators}.

Using the same methods we can also compute the conductivity matrix $\sigma$, associated with the small frequency limit of the retarded current-current correlator $G^{i,j}_R$, $\sigma = \frac{i}{\omega} G^{i,j}_R$,
\begin{multline}
\label{E:sigmafinal}
	\sigma  
		= -\frac{N^2 \pi^2 T}{V_5 \gamma} 
		\begin{pmatrix} 
		\frac{1}{8} + s_0  	&  0  &  0  \\
		\left( \frac{i \pi T \Delta}{\omega} - s_1 \right) \sin (k x) & \frac{1}{8} + 2 s_0 - s_2 \cos(2 k x) & s_2 \sin (2 k x)  \\
		\left( \frac{i \pi T \Delta}{\omega} - s_1 \right) \cos (k x) & s_2 \sin(2 k x) & \frac{1}{8}  + 2 s_0 + s_2 \cos(2 k x)
		\end{pmatrix}
		\\
		+\mathcal{O}(\omega^2, \gamma^{-3})
\end{multline}
with $T$ the temperature, $s_i$ dimensionless functions, $\Delta$ the order parameter for the modulated phase, $N$  the rank of the gauge group and $V_5$ is a model dependent number. When $\Delta = 0$, the dimensionless functions $s_i$ vanish and $\sigma$ reduces to the conductivity of a homogeneous isotropic phase dual to a probe AdS-RN black brane \cite{Son:2006em, Herzog:2010vz}. In obtaining \eqref{E:sigmafinal} we have used the canonical  relation
\[
	G_5 = \frac{V_5}{2 \pi^2 N^2}
\]
between Newtons constant in the bulk and the rank of the gauge group $N$ of the boundary theory. The numerical value of $V_5$ is theory dependent, see e.g., \cite{Gubser:1998vd}. For $\mathcal{N}=4$ super Yang-Mills theory, $V_5=\pi^3$. 

Similarly, the current-stress and stress-current correlators at low frequency are given by
\begin{align}
\begin{split}
\label{E:mixedfinal}
	G_R^{xy,i} & = \frac{N^2 \pi^4 T^3 i \omega}{2 V_5 \gamma^2} \left(\frac{T}{k} \cos(k x) u_0,\,0,\,u_1\right)  +  \mathcal{O}(\omega^2 ,\gamma^{-4}) \\
	G_R^{i,xy} &= \frac{N^2 \pi^3 T^3}{4 V_5 \gamma}  \left( \Delta \cos(k x),\,\frac{i\omega}{T} \sin(2 k x)u_2 ,\, -\frac{i \omega}{T} u_3 + \frac{i\omega}{T} \cos(2 k x) u_2 \right)  +  \mathcal{O}(\omega^2 ,\gamma^{-4}) \\
	G_R^{i,yz} &= \frac{N^2 \pi^4 T^3}{4 V_5 \gamma}  \left( 0,\,-\sin( k x) g + i \frac{\omega}{T} \sin(3 k x) u_5,\, \cos( k x)g + i \frac{\omega}{T} \cos(3 k x) u_5 \right)  +  \mathcal{O}(\omega^2 ,\gamma^{-4}) \,,
\end{split}
\end{align}
with
\begin{equation}
	g = -\frac{\Delta}{2} + i \frac{\omega}{T} u_4 \,.
\end{equation}
\end{subequations}
As before, the $u_i$ are dimensionless functions. A full discussion of \eqref{E:finalresult} is left to section \ref{S:Discussion}.

This work is organized as follows: In section \ref{S:Instability} we review the instability discussed in \cite{Nakamura:2009tf} and elaborated on in \cite{Donos:2012wi}. This introductory section also sets the notation for the rest of the paper. In section \ref{S:Background} we construct the spatially modulated black brane solution perturbatively in $\gamma^{-1}$ and $\Delta$ in terms of Heun functions. The details of the computation of the correlation functions \eqref{E:finalresult} can be found in Section \ref{S:Correlators} and Appendix \ref{A:Correlators}. We discuss our results in section \ref{S:Discussion}.

\section{A spatially modulated phase from a $U(1)^3$ anomaly}
\label{S:Instability}

The black brane solution to the equations of motion following from \eqref{E:bulkaction} is given by
\begin{subequations}
\label{E:AdSRN}
\begin{equation}
	ds^2 = -g(r)dt^2 + \frac{dr^2}{g(r)} + r^2 \left(dx^2 + dy^2 + dz^2\right)
	\qquad
	A = a dt\,,
\end{equation}
where
\begin{equation}
	g = r^2 - \frac{r_+^4}{r^2} + \frac{\mu^2}{3} \left(\frac{r_+^4}{r^4} - \frac{r_+^2}{r^2}\right)
	\qquad
	a= \mu \left(1- \frac{r_+^2}{r^2}\right)\,.
\end{equation}
\end{subequations}
The asymptotically AdS boundary is located at $r\to\infty$ and the event horizon is at $r=r_+$. The standard AdS/CFT dictionary \cite{Witten:1998qj} implies that the solution \eqref{E:AdSRN} corresponds to a state of the field theory with chemical potential $\mu$ and temperature
\begin{equation}
	T = \frac{r_+}{\pi} - \frac{\mu^2}{6 \pi r_+}\,.
\end{equation}

As discussed in \cite{Donos:2012wi}, the equations of motion following from \eqref{E:bulkaction} also admit helical black brane solutions with partially broken translation invariance, i.e., the gauge field and metric are invariant under $\partial_{y}$ and $\partial_{z}$ but are not translationally invariant in the $x$ direction. Instead they are invariant under a combined translation in the $x$ direction and rotation in the $(y,\,z)$ plane which is generated by $\xi = \partial_{x} - k \left(y \partial_{z} - z \partial_{y}\right)$. The ansatz for the solution respecting this helical symmetry is given by \cite{Donos:2012wi}
\begin{align}
\label{E:helicalansatz}
\begin{split}
	ds^2 &= - g f^2 dt^2 + \frac{dr^2}{g} + h^2 dx^2 + r^2 e^{2v} \left(\omega_y + Q dt\right)^2 + r^2 e^{-2v} \omega_z^2
	\\
	A &= a dt+b \omega_y
\end{split}
\end{align}
where
\begin{equation}
\label{E:omegayz}
	\omega_y = \cos(kx)dy-\sin(kx)dz\,,
	\qquad
	\omega_z = \cos(kx) dy+ \sin(kx) dz\,.
\end{equation}
Note the $\mathbb{Z}_2$ symmetry, $x\to-x$ and $z\to-z$.

The equations of motion associated with \eqref{E:helicalansatz} are somewhat long-winded and we will refrain from writing them down explicitly. Setting $v=Q=b=0$ and $h=r$ we recover the charged black brane solution in \eqref{E:AdSRN} which has enhanced translational and rotational symmetry in the $x$, $y$ and $z$ directions. Requiring that the space-time be asymptotically AdS implies that the near boundary (large $r$) expansion of the metric and gauge field take the form
\begin{align}
\begin{split}
	g = & r^{2}\left(1-\frac{M}{r^{4}}+\ldots\right) 
	\qquad
	f =  1-\frac{c_{h}}{r^{4}}+\ldots
	\qquad
	h =  r\left(1+\frac{c_{h}}{r^{4}}+\ldots\right)
	\\
	v = & \frac{c_{v}}{r^{4}}+\ldots
	\qquad
	Q =  \frac{c_{Q}}{r^{4}}+\ldots
	\qquad
	a =  \mu+\frac{q}{r^{2}}+\ldots
	\qquad
	b =  \frac{c_{b}}{r^{2}}+\ldots\,.
\end{split}
\end{align}
Requiring that the space-time and gauge field be regular at the (outermost) event horizon located at $r=r_+$ implies that
\begin{align}
\begin{split}
	g & = g_{+}(r-r_{+}) +\ldots
	\qquad
	f = f_{+}+\ldots
	\qquad
	h = h_{+}+\ldots
	\qquad
	v  = v_{+}+\ldots
	\\
	Q & = Q_{+}(r-r_{+})+\ldots  
	\qquad
	a  = a_{+}(r-r_{+})+\ldots
	\qquad
	b  = b_{+}+\ldots\,.
\end{split}
\end{align}

The standard AdS/CFT dictionary implies that the helical black holes \eqref{E:helicalansatz} are dual to field theory configurations with spontaneously generated helical symmetry where the stress tensor and current are given by
\begin{equation}
	16 \pi G_5 \left\langle T_{\mu\nu}\right\rangle =\left(\begin{array}{cccc}
		3M+8c_{h} & 0 & 4c_{Q}\cos\left(kx\right) & -4c_{Q}\sin\left(kx\right)\\
		0 & M+8c_{h} & 0 & 0\\
		4c_{Q}\cos\left(kx\right) & 0 & M+8c_{v}\cos\left(2kx\right) & -8c_{v}\sin\left(2kx\right)\\
		-4c_{Q}\sin\left(kx\right) & 0 & -8c_{v}\sin\left(2kx\right) & M-8c_{v}\cos\left(2kx\right)
	\end{array}\right)\,,
\end{equation}
and
\begin{equation}
\label{E:Jvev}
	\left\langle J_{\mu}\right\rangle =\left(-2q,0,-2c_{b}\cos\left(kx\right),2c_{b}\sin\left(kx\right)\right)\,.
\end{equation}
The temperature of this configuration is given by
\begin{equation}
\label{E:Temperature}
	T = \frac{g_+ f_+}{4\pi}
\end{equation}
and the chemical potential is $\mu$. We will find that the value 
\begin{equation}
\label{E:entropy}
	s= \frac{r_+^2 h_+}{4 G_5}
\end{equation}
for the entropy density and
\begin{equation}
\label{E:FreeEnergy}
	16 \pi G_5 F = - M
\end{equation}
for the free energy as computed in \cite{Donos:2012wi}  to be particularly useful in what follows.

A helical solution of the type \eqref{E:helicalansatz} can exist only when $\gamma$ is above a critical value $\gamma_c\sim 0.1448>1/4\sqrt{3}$ and below a critical temperature $T_c(\gamma,k)$ \cite{Nakamura:2009tf}. Above the critical temperature the solution to the equations of motion is the AdS black hole described in \eqref{E:AdSRN}. Below the critical temperature there are two solutions to the equations of motion: the charged black hole solution given by \eqref{E:AdSRN} and a helical solution for which $b$, $Q$ and $\alpha$ obtain non trivial values. The helical solution has a lower free energy and is therefore the preferred phase.

It is  convenient to study the phase transition to the helical  phase  in the probe limit where we take $\gamma$ to be very large and $A/\gamma$ to be finite. Since the stress tensor is quadratic in the gauge fields, the equations of motion for the gauge field and metric decouple at every order in $1/\gamma$. To leading order in $1/\gamma$ the equations of motion reduce to the Einstein equations in the absence of matter whose solution is the AdS black hole for which
\begin{equation}
\label{E:AdSSS}
	g = r^2 - \frac{r_+^4}{r^2}
	\qquad
	f=1
	\qquad
	h=r
\end{equation}
and the remaining metric components vanish. The solution \eqref{E:AdSSS} can be obtained by taking the $\mu\to0$ limit of \eqref{E:AdSRN}.  In what follows we will switch to a coordinate system where 
\begin{equation}
	u= r^2 / r_+^2\,.
\end{equation}

At subleading order in $1/\gamma$ we find the following set of coupled equations for the components of the gauge field $A$,
\begin{equation}
\label{E:abequations}
	{a}''(u) = \frac{4 k}{r_+^2} b(u) b'(u)
	\qquad
	((1-u^2)b'(u))' - \frac{k^2}{4 r_+^2 u} b(u)  - \frac{4 k}{r_+^2} b(u) a(u) = 0
\end{equation}
where a $'$ denotes a derivative with respect to $u$.
The boundary conditions we impose on \eqref{E:abequations} are that $a$ vanish at the black hole horizon $r=r_+$, that $b=0$ at the asymptotically AdS boundary $r \to \infty$ and that $b$ is finite at the horizon. The chemical potential of the boundary theory can be read off of the boundary value of $a$, $a(\infty) = \mu$. Inserting the solution for $a$ into the equation for $b$ one obtains a non-linear integro-differential equation for $b$. 

The trivial solution to \eqref{E:abequations} is 
\begin{equation}
\label{E:aleading}
	b=0 
	\qquad
	a = \mu \left( 1 - \frac{r_+^2}{r^2} \right)
\end{equation}
reproducing the gauge potential for the charged black hole solution \eqref{E:AdSRN}. Since the equations are non linear, there  exist other, non trivial solutions to the equations of motion for $b$. To get a handle on these non trivial solutions  we look for instabilities of  \eqref{E:aleading} to small perturbations. The equation of motion for perturbations $\beta$ of $b$ around \eqref{E:aleading} is
\begin{equation}
\label{E:eigenequation}
	((1-u^2) \beta ')' + 4 \hat{\mu} \hat{k} \,\beta = \frac{\hat{k}^2}{4u} \beta\,,
\end{equation} 
where we have defined
\begin{equation}
	\hat{k} = \frac{k}{r_+} 
	\qquad
	\hat{\mu} = \frac{\mu}{r_+}
\end{equation}
and we require that $\beta(0) = 0$ and $\beta(1) = \hbox{finite}$ (that the field $b$ is not sourced and is finite at the horizon). Equation \eqref{E:eigenequation} is a Heun equation which has been extensively studied in the literature  \cite{0198596952}. The eigenvalue problem we wish to solve is equivalent to the problem of finding Heun functions. For certain special values of $\hat{k}$ and $\hat{\mu}$ the Heun functions reduce to Heun polynomials. For instance, for $\hat{\mu}=3/\sqrt{32\sqrt{2}}$ and $\hat{k} = \sqrt{8\sqrt{2}}$ we find that $\beta=u+\sqrt{2}u^2$ solves the equations of motion.  

To get a better understanding of the solution to \eqref{E:eigenequation} we proceed as follows. For each value of $\theta= \hat{\mu}\hat{k}$, let us think of the Sturm-Liouville system in \eqref{E:eigenequation} as an eigenvalue problem where the left hand side of \eqref{E:eigenequation} defines the Sturm-Liouville operator $S_{\theta}$. Sturm Liouville theory guarantees that that (for each $\theta$) the solutions to
\begin{equation}
\label{E:eigenexplicit}
	S_{\theta} \beta_n  = -\frac{\lambda_n}{u} \beta_n
\end{equation}
are characterized by an infinite set of real $\lambda_n$ where $\lambda_n \to \infty$ for large $n$. Moreover, if we order the eigenvalues such that 
\begin{equation}
\label{E:lambdaorder}
	\lambda_0 < \lambda_1 < \ldots
\end{equation}
then the eigenfunction $\beta_n$ has exactly $n$ zeros in the interval $0<u<1$.  Since $\hat{k}^2$ is positive, physical solutions to \eqref{E:eigenequation} exist whenever $S_{\theta}$ has negative eigenvalues, viz. $\hat{k}_n =2 \sqrt{-\lambda_n(\theta)}$ and $\hat{\mu} = \theta/2\sqrt{-\lambda_n(\theta)}$.

\begin{figure}[hbt]
\centering
	\includegraphics[width =\linewidth]{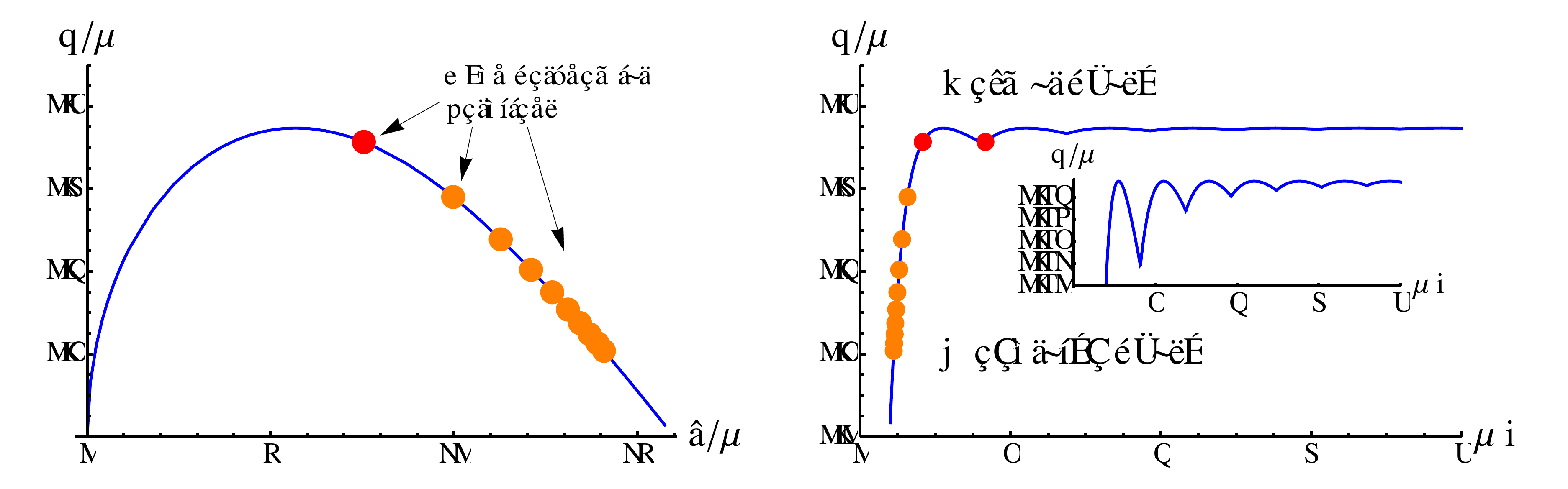}
\caption{\label{F:critcurve}  
(Left) A plot of the critical curve where the AdS black hole becomes unstable. Several points on the critical curve which may be described by Heun polynomials are shown in red and orange. Infinitely many other such points exist. (Right) The phase diagram of the helical phase for a compact $x$ direction, $x \sim x+L$.  The boundary of the normal phase is fixed by the highest temperature for which there exists an integer $n$ and a pair $(T/\mu, 2 \pi n/L \mu)$ on the critical curve on the left.
}
\end{figure}

To visualize the region in $k/\mu$, $T/\mu = r_+/\pi\mu$ space where solutions to \eqref{E:eigenequation} exist, consider the following. For each $\theta$ consider all $\lambda_n(\theta)<0$ where $n=0,\ldots, N$. For each such set of $\lambda_n$'s we have the set of points
\begin{equation}
	\frac{k_n}{\mu_n} = \frac{\hat{k}_n}{\hat{\mu}_n} = 4\frac{ |\lambda_n(\theta)| }{\theta}
	\qquad
	\frac{T}{\mu_n} = \frac{1}{\pi\hat{\mu}_n} = 2\frac{\sqrt{|\lambda_n(\theta)|}}{ \pi \theta}\,,
\end{equation}
which satisfy $T_0 > T_1 > \ldots > T_{N}$. Since $\lambda_n(\theta)$ are continuous in $\theta$ and satisfy \eqref{E:lambdaorder}, by varying the value of $\theta$ we generate sets of non intersecting curves in the $T/\mu$, $k/\mu$ plane which terminate only at $T=0$. The curve associated with the smallest eigenvalue will have the highest temperature for any $k$ and its associated eigenfunction will have no zeros on the interval $0 < u < 1$. We will refer to this curve as the critical curve. A numerical evaluation of the critical curve can be found in the left panel of  figure \ref{F:critcurve} (blue).  Some points on the critical curve (associated with Heun polynomial solutions to \eqref{E:eigenequation}) can be obtained analytically as we show in appendix \ref{A:analyticzeromode}. 

The instability associated with the critical curve implies the existence of a new phase whose stability properties can be inferred by computing its free energy. Such an endeavor was carried out in \cite{Donos:2012wi}. Working in the probe approximation this task becomes somewhat simpler and can be handled analytically to an extent. We find that the isotropic phase has a higher free energy and is therefore less stable.

If the volume of the spatial direction specified by the $x$ coordinate is infinite then all values of $k$ are allowed. The critical temperature at which the new, modulated, phase will appear and the momentum associated with the modulated phase can be read off of the extremum of the critical curve in figure \ref{F:critcurve}. If the $x$ direction is compactified then the momentum $k$ becomes discrete $k = 2 \pi n/L$. Hence, for each value of $\mu L$ there are only finitely many allowed momenta. By fixing $\mu L$ we may force the system to undergo a phase transition at a point on the critical curve other than its extremum. For instance, we may tune $\mu L$ so that the phase transition occurs at a point described by a Heun polynomial. The phase diagram for the helical phase can be found in the right panel of figure \ref{F:critcurve}. 

\section{Solving the equations of motion in the probe limit}
\label{S:Background}
We have seen in the previous section that in the probe limit, when $\gamma$ is very large, the equations of motion for the matter fields and the metric decouple. Consider the expansion 
\begin{align}
\begin{split}
\label{E:probelimit}
	g &= \sum_{n=0}^{\infty} \gamma^{2n} g_{2n}(r) \qquad
	h = \sum_{n=0}^{\infty} \gamma^{2n} h_{2n}(r) \qquad
	f = \sum_{n=0}^{\infty} \gamma^{2n} f_{2n}(r) \qquad
	v = \sum_{n=0}^{\infty} \gamma^{2n} v_{2n}(r) \\
	Q &= \sum_{n=0}^{\infty} \gamma^{2n} Q_{2n}(r) \qquad
	a = \sum_{n=0}^{\infty} \gamma^{2n+1} a_{2n+1}(r) \qquad
	b = \sum_{n=0}^{\infty} \gamma^{2n+1} b_{2n+1}(r) \,.
\end{split}
\end{align}
At order $\gamma^{0}$ the equations of motion are the Einstein equations in the absence of matter. At order $\gamma^{-1}$ the equations describe a Maxwell-Chern-Simons theory in an AdS black brane background. The order $\gamma^{-2}$ equations describe the backreaction of the metric to the matter fields. We will now solve the equations of motion order by order in $\gamma$.

\subsection{The background metric}
At order $1/\gamma^0$ the equations of motion reduce to the Einstein equations in the absence of matter. Imposing the boundary conditions  $g_0 \to r_+^2/u$, $f_0 \to 1$ and  $h_0 \to \sqrt{r_+^2/u}$ we find that the solution to the equations of motion is the AdS Schwarzschild black hole given in equations \eqref{E:AdSSS}, 
\begin{equation}
	g_0 = r_+^2 \left(u^{-1} - u\right)
	\qquad
	f_0=1
	\qquad
	h_0 = \sqrt{\frac{r_+^2}{u}} \,.
\end{equation}
At this order in $\gamma$ we find that \eqref{E:Temperature}-\eqref{E:FreeEnergy} reduce to 
\begin{equation}
	r_+ = \pi T
	\qquad
	s=\frac{N^2 \pi^5 T^3}{2 V_5}
	\qquad
	F = -\frac{N^2 \pi^5}{8 V_5} T^4\,.
\end{equation}

\subsection{The leading term for the gauge field}
At subleading order in $1/\gamma$ we find the equations of motion \eqref{E:abequations} for the gauge field
\begin{equation}
\label{E:abequationsV2}
	\hat{a}_1'' =4 \hat{k} \hat{b}_1  \hat{b}'_1
	\qquad
	\left((1-u^2) \hat{b}_1'\right)' - \frac{\hat{k}^2 }{4 u} \hat{b}_1 = 4 \hat{k} \hat{b}_1 \hat{a}'_1\,,
\end{equation}
where we have used
\begin{equation}
	\hat{a}_1 = \frac{a_1}{r_+}
	\qquad
	\hat{b}_1 = \frac{b_1}{r_+}
	\qquad
	\hat{k} = \frac{k}{r_+} 
	\qquad
	\hat{\mu} = \frac{\mu}{r_+}\,.
\end{equation}
In what follows hatted quantities will always demote a dimensionless version of their unhatted counterparts, obtained via multiplication by an appropriate factor of $r_+$.
The boundary conditions we impose on \eqref{E:abequationsV2} are that  at the asymptotically AdS boundary ($u=0$) we have $a(0) = \mu$ which defines the chemical potential at the boundary and that $b(0)=0$ which implies that the spatial component of the current is not sourced. At the horizon ($u=1$) we impose that $a_1(1)=0$ and that $b_1(1) = \hbox{finite}$. Equations \eqref{E:abequationsV2} can be easily solved using standard numerical integration schemes. We find that a solution exists anywhere inside the critical curve in the $k/\mu$, $T/\mu$ plane described in the previous section. 
As suggested earlier, one may solve for $a_1$ in terms of $b_1$,
\begin{equation}
\label{E:a1}
	\hat{a}_1 = \left(\hat{\mu} + 2 \hat{k} \int_0^1 \hat{b}_1^2(t)\,dt \right) (1-u) - 2 \hat{k} \int_u^1 \hat{b}_1^2(t) \,dt\,.
\end{equation}
Inserting \eqref{E:a1} into the equation of motion for $b_1$ one obtains a non linear integro-differential equation which can not be solved analytically.

\subsection{The backreaction of the metric}

At second order in $1/\gamma$ the equations of motion for the backreaction of the metric may be arranged into a nested set of linear equations,
\begin{align}
\begin{split}
\label{E:metricbackreaction}
	\left(\frac{(1-u^2)\left( \sqrt{u} \hat{h}_2 \right)'}{u}  \right) ' =& -\frac{\hat{k}^2}{16 u}   \hat{b}^2_1  + \frac{1}{4} (1-u^2) \left({\hat{b}_1^{\prime}}\right)^2 \\
	\left(\frac{1}{u}\hat{g}_2 - \frac{4}{3}\sqrt{u}  \hat{h}_2\right)' = & 
	\frac{1}{3} \left(\hat{\mu} + 2 \hat{k} \int_0^1 \hat{b}_1^2(t) dt - 2 \hat{k}   \hat{b}^2_1  \right)^2
	+ \frac{2}{3} (1-u^2) \left( \hat{b}_1' \right)^2	\\
	f_2' + \frac{1}{3}  \frac{3+u^2}{1-u^2} \left(\sqrt{u} \hat{h}_2 \right)' = & \frac{u}{2} \left( \frac{\hat{k}^2 \hat{b}^2_1}{12 (1-u^2)}  -u \left( \hat{b}_1' \right)^2 \right)\\
	\left(\frac{Q_2'}{u}\right)' - \frac{\hat{k}^2}{4 u^2 (1-u^2)} Q_2 = &  \left[ \hat{\mu} + 2 \hat{k} \left( \int_0^1 \hat{b}_1^2 dt - \hat{b}_1^2 \right) \right] \hat{b}_1' \\
	\left(\frac{(1-u^2)}{u} v_2' \right)' - \frac{\hat{k}^2}{u^2} v_2 = & \frac{\hat{k}^2}{16 u}  \hat{b}_1^2 - \frac{1-u^2}{4} \left( \hat{b}_1' \right)^2\,
\end{split}
\end{align}
where
\begin{equation}
	\hat{h}_2 = \frac{{h}_2}{r_+}
	\qquad
	\hat{g}_2 = \frac{g_2}{r_+^2}
	\qquad
\end{equation}
(and if we were to use hatted versions of $f_2$, $Q_2$ and $v_2$ we would have found that $\hat{f}_2 = f_2$, $\hat{Q}_2 = Q_2$ and $\hat{v}_2 = v_2$).
The boundary conditions we impose on \eqref{E:metricbackreaction} is that the horizon is at $u=1$, viz., $Q_2(1)=\hat{g}_2(1) = 0$, and $\hat{h}_2(1)$ and $v_2(1)$ are finite. Requiring that the geometry is asymptotically AdS implies that $\lim_{u\to0} \sqrt{u}\hat{h}_2 = 0$, $\lim_{u\to0} u\hat{g}_2=0$, $\hat{f}_2(0)=0$, $\hat{Q}_2(0)=0$ and $\hat{v}_2(0)=0$.

The first three equations in \eqref{E:metricbackreaction} can be solved implicitly in terms of integrals of $b_1$. The resulting expressions are somewhat long-winded and we will not present them here.
The last two equations in \eqref{E:metricbackreaction} can be solved in terms of the homogenous solutions to the $Q_2$ and $v_2$ equations. For instance, let $\mathfrak{v}_{1}$ and $\mathfrak{v}_{2}$ satisfy
\begin{equation}
	\left(\frac{(1-u^2)}{u} \mathfrak{v}_i' \right)' - \frac{\hat{k}^2}{u^2} \mathfrak{v}_i = 0 \,.
\label{E:vHomogenousEquation}
\end{equation}
From the asymptotic behaviour of the general solution to \eqref{E:vHomogenousEquation} near $u = 0$ and $u = 1$, the solutions $\mathfrak{v}_1, \mathfrak{v}_2$ can be chosen to be of the form
\begin{equation}
	\mathfrak{v}_2 = u^2 + \mathcal{O}(u^3)
	\qquad
	\mathfrak{v}_1 = 1 - \hat{k}^2 u + \mathcal{O}(u^2)
\end{equation}
near the boundary ($u=0$) with $\mathfrak{v}_1$ finite at the horizon ($u=1$) while $\mathfrak{v}_2$ is allowed to diverge there.  Then
\begin{equation}
		v_{2}  =  \frac{1}{2} \mathfrak{v}_1 \int_0^u \mathfrak{v}_2(t) S_{v}(t) dt + \frac{1}{2} \mathfrak{v}_2 \int_u^1 \mathfrak{v}_1(t) S_{v}(t) dt 
\end{equation}
where
\begin{equation}
	S_{v} =  \frac{\hat{k}^2}{16 u} \hat{b}_1^2 - \frac{1-u^2}{4} \left(\hat{b}_1^{\prime}\right)^2\,.
\end{equation}

A straightforward though somewhat tedious computation also allows us to compute the near boundary asymptotics of $h_2$. We find that
\begin{equation}
\label{E:chval}
	\frac{c_h}{r_+^4}  = -\frac{1}{2} \hat{\mu} \hat{k} \int_0^1\hat{b}_1^2(t) \, dt - \hat{k}^2 \left( \int_0^1 \hat{b}_1^2(t) \, dt \right)^2 
	+ \frac{1}{16} \hat{k}^2 \int_0^1  \frac{\hat{b}_1^2(t)}{t}\, dt + \hat{k}^2 \int_0^1 \hat{b}_1^4(t) \, dt  \,.
\end{equation}
In \cite{Donos:2013cka} it was shown that for fixed $T/\mu$, the free energy is minimized at values of $\hat{k}/\hat{\mu}$ which satisfy $c_h=0$. 	

\subsection{Perturbative analysis}
\label{S:PerturbativeBgd}
In order to have explicit expressions for the response of the metric to the gauge field we need, at the very least, an explicit expression for $b_1(u)$.
If we restrict ourselves to the behavior of $b_1$ near the phase transition we may carry out a perturbative expansion near the critical curve. We will find this construction very useful in what follows. For each value of $\hat{k}$ and $\hat{\mu}$ inside the critical curve there is an associated value for the expectation value of the spatial part of the current, $c_b$, c.f., equation \eqref{E:Jvev}, which also serves as an order parameter for the helical phase. We find it convenient to work perturbatively in this order parameter, $\Delta = c_b/r_+^3$. Let us denote
\begin{equation}
\label{E:abexpansion}
	b_1 = \sum_{i=0}^{\infty} \Delta^{2i+1} b_{1,2i+1}
	\qquad
	a_1 = \sum_{i=0}^{\infty} \Delta^{2i} a_{1,2i}\,.
\end{equation}
and also
\begin{equation}
\label{E:kmuexpansion}
	\hat{k} = \hat{k}_ 0  + \sum_{i=0}^{\infty} \Delta^{2n} \widehat{\delta {k}}_{2n}
	\qquad
	\hat{\mu} = \hat{\mu}  + \sum_{i=0}^{\infty} \Delta^{2n} \widehat{\delta {\mu}}_{2n}
\end{equation}	
where $\hat{k}_0$ and $\hat{\mu}_0$ are points on the critical curve. 

Working perturbatively to order $\Delta^2$ we find that
\begin{equation}
	\hat{a}_1 = \hat{\mu}_0  (1-u) + \Delta^2 \left[
		\left( \widehat{\delta  {\mu}}_2  + 2 \hat{k}  \int_0^1   \hat{b}_{1,1}^2(t) \, dt \right)(1-u) - 2 \hat{k}_0 \int_u^1   \hat{b}_{1,1}^2(t) \, dt 
		\right]
		+ \mathcal{O}(\Delta^4)
\end{equation}
where $\hat{b}_{1,1}$ satisfies the linear homogenous equation
\begin{equation}
\label{E:Ldef}
	L\, \hat{b}_{1,1} \equiv ((1-u^2) \hat{b}_{1,1} ')' + 4 \hat{\mu}_0 \hat{k}_0 \,\hat{b}_{1,1} - \frac{\hat{k}_0^2}{4 u} \hat{b}_{1,1} = 0
\end{equation}
with boundary conditions such that $\hat{b}_{1,1}  = u + \mathcal{O}(u^2)$ near the boundary and that $\hat{b}_{1,1}$ is finite at the horizon. 
Equation \eqref{E:Ldef} is identical to the Heun equation \eqref{E:eigenequation} which we have studied in the previous section. The particular solutions $\hat{b}_{1,1}$ are Heun functions. For particular values of $\hat{k}_0$ and $\hat{\mu}_0$ the Heun functions reduce to Heun polynomials. For instance
\begin{subequations}
\label{E:lowest}
\begin{equation}
	\hat{b}_{1,1} = u + \sqrt{2}u^2
\end{equation}
together with 
\begin{equation}
	\hat{k}_0  = \sqrt{8\sqrt{2}}
	\qquad
	\hat{\mu}_0 =  \frac{3}{\sqrt{32\sqrt{2}}}
\end{equation}
\end{subequations}
is a solution to \eqref{E:Ldef}. Since this solution does not correspond to the maximum of the critical curve \ref{F:critcurve}, in order for it to be physically relevant we need to choose an appropriately sized box so that the only available momentum is $\hat{k}_0$. A classification of Heun polynomials solving \eqref{E:Ldef} can be found in appendix \ref{A:analyticzeromode}.

At third order in $\Delta$ we obtain the equation
\begin{equation}
\label{E:b13eom}
	L\, \hat{b}_{1,3}  = -\left( 4 \widehat{ \delta \mu}_2 \hat{k}_0 + \widehat{\delta k}_2 \left( 4 \hat{\mu}_0 - \frac{\hat{k}_0}{2 u} \right) \right) \hat{b}_{1,1}  - 8 \hat{k}_0^2 \hat{b}_{1,1}  \left( \int_0^1 \hat{b}_{1,1}^2(t)\,dt -  \hat{b}_{1,1}^2 \right)\,,
\end{equation}
with the boundary condition that ${b}_{1,3} = \mathcal{O}(u^2)$ near the boundary and that $b_{1,3}$ is finite at the horizon. Consider
\begin{equation}
\label{E:mkconstraint}
	\int_0^1 b_{1,1} L \,b_{1,3} dt = \int_0^1 b_{1,1} L \,b_{1,3} - b_{1,3} L \,b_{1,1} dt= 0
\end{equation}
where the last equality follows from inserting the explicit form for $L$ defined in \eqref{E:Ldef} and rewriting the resulting expression as a total derivative. Equation \eqref{E:mkconstraint} may be thought of as a constraint on the possible values of $\widehat{\delta {\mu}}_2$ and $\widehat{\delta {k}}_2$ so that $\hat{\mu}$ and $\hat{k}$ best approximate their true values at small $\Delta$, c.f., equation \eqref{E:kmuexpansion}. Carrying out the integral on the left hand side of \eqref{E:mkconstraint} explicitly and using \eqref{E:b13eom} we find
\begin{multline}
\label{E:deltacurve}
	4 \widehat{\delta {\mu}}_2 \hat{k}_0 \int_0^1 \hat{b}_{1,1}^2(t)\,dt+ \widehat{\delta {k}}_2 \left(4 \hat{\mu}_0 \int_0^1 \hat{b}_{1,1}^2(t)\,dt - \frac{1}{2} \hat{k}_0 \int_0^1 \frac{\hat{b}_{1,1}^2(t)}{t}\,dt \right) \\
	= 8 \hat{k}_0^2 \left( \int_0^1 \hat{b}_{1,1}^4(t)\,dt - \left( \int_0^1 \hat{b}_{1,1}^2(t)\,dt \right)^2 \right) \,.
\end{multline}
Thus, starting from the top of the critical curve located at $\hat{k}_0$, $\hat{\mu}_0$ we may probe a small $\Delta$ region around this extremum located on the curve \eqref{E:deltacurve}. The particular values of $\hat{\mu}$ and $\hat{k}$ for which the free energy is minimized to order $\Delta$ are given by \eqref{E:deltacurve} together with \eqref{E:chval}.

With the $\mathcal{O}(1/\gamma)$ solutions at hand we may also expand the $\mathcal{O}(1/\gamma^2)$ solution of the backreacted metric order by order in $\Delta$,
\begin{align}
\begin{split}
	h_2 &= \sum_{i=0} \Delta^{2i} h_{2,2i}
	\qquad
	g_2 = \sum_{i=0} \Delta^{2i} g_{2,2i}
	\qquad
	f_2 = \sum_{i=0} \Delta^{2i} f_{2,2i}
	\\
	v_2 &= \sum_{i=0} \Delta^{2i} v_{2,2i}
	\qquad
	Q_2 = \sum_{i=0} \Delta^{2i+1} Q_{2,2i+1}
\end{split}
\end{align}
An explicit computation yields
\begin{align}
\begin{split}
	h_{2,0} &=0
	\qquad
	g_{2,0} = -\frac{\hat{\mu}_0}{3} u (1-u) 
	\qquad
	f_{2,0} = 0
	\qquad
	v_{2,0} = 0
\end{split}
\end{align}
and
\begin{align}
\begin{split}
	h_{2,2} =& -\frac{1}{8\sqrt{u}}\ln(1-u^2) \left(\frac{\hat{k}_0^2}{4} \int_u^1 \frac{\hat{b}_{1,1}^2(t)}{t} \,dt - \int_u^1 (1-t^2) \left(\hat{b}_{1,1}'(t)\right)^2 dt \right) \\
	& -\frac{1}{8\sqrt{u}} \left(\frac{\hat{k}_0^2}{4} \int_0^u \frac{\hat{b}_{1,1}^2(t)}{t}\ln(1-t^2) dt -  \int_0^u (1-t^2)\ln(1-t^2) \hat{b}_{1,1}'(t)dt \right) \\
	g_{2,2} =& \frac{1}{3} u \int_u^1 \frac{s}{1-s^2} \int_s^1 (1-t^2)\left(\hat{b}_{1,1}^{\prime}(t)\right)^2\, dt \,ds 
		- \frac{1}{12}u\hat{k}_0^2 \int_u^1 \frac{s}{1-s^2} \int_s^1 \frac{\hat{b}_{1,1}^2(t)}{t} \,dt \,ds \\
		&- \frac{2}{3}(1-u)u \,\widehat{\delta\mu}_2\, \hat{\mu}_0 - \frac{4}{3} u \left((1-u) \int_0^1 \hat{b}_{1,1}^2(t)\,dt- \int_u^1 \hat{b}_{1,1}^2(t)\,dt \right)\hat{k}_0 \hat{\mu}_0 \\
		&- \frac{2}{3} u \int_u^1 (1-s^2) \left(\hat{b}_{1,1}\right)' ds \\
	f_{2,2} = & \frac{1}{12} \int_0^u \frac{(3+s^2)s}{(1-s^2)^2} \int_s^1 (1-t^2) \left(\hat{b}_{1,1}'(t)\right)^2 \,dt \,ds + \frac{1}{24} \hat{k}_0^2 \int_0^u \frac{s}{1-s^2} \left(\hat{b}_{1,1}^2(s)-\hat{b}_{1,1}^2(1)\right) ds \\
	&- \frac{1}{48} \hat{k}_0^2 \int_0^u \frac{(3+s^2)s}{(1-s^2)^2} \left( \int_u^1 \frac{\hat{b}_{1,1}^2(t)}{t}\,dt + \hat{b}^2_{1,1}(1)(s-1) \right) ds -\frac{1}{2} \int_0^u s^2 \left(\hat{b}_{1,1}'(s)\right)^2 ds \\
	v_{2,2} & =  \frac{1}{2} \mathfrak{v}_1 \int_0^u \mathfrak{v}_2(t) S_{v}(t) dt - \frac{1}{2} \mathfrak{v}_2 \int_u^1 \mathfrak{v}_1(t) S_{v}(t) dt 
\end{split}
\end{align}
where
\begin{equation}
	S_{v} =  \frac{\hat{k}_0^2}{16 u} \hat{b}_{1,1}^2 - \frac{1-u^2}{4} (\hat{b}_{1,1}^{\prime})^2\,.
\end{equation}
We have not included similar looking expressions for $Q_{2,1}$.

\section{Correlators in the helical phase}
\label{S:Correlators}

With the modulated background at our disposal we are now ready to compute correlators in this background. Our strategy for computing correlators is to add a local source term to the operator whose correlator we compute, and calculate its response to the source. For instance, if we add a time dependent metric perturbation $h_{lm}(t)$ we find that the response of the stress tensor to the perturbation is given by
\begin{equation}
\label{E:deltaT}
	\delta T_{\mu\nu} = -\frac{1}{2} \int G_R^{\mu\nu,\rho\sigma}(t,\vec{x};t',\vec{x}')h_{\rho\sigma}(t') dt' d^3x'  + \mathcal{O}(h^2)
\end{equation}
where
\begin{equation}
	G_R^{\mu\nu,\rho\sigma} = - i \theta(t-t') [T^{\mu\nu}(t,\vec{x}),\,T^{\rho\sigma}(t',\vec{x}')]
\end{equation}
is the retarded Green function. We have omitted contact terms from \eqref{E:deltaT}.

Let us define the Fourier transformed Greens function:
\begin{equation}
	G_R^{\mu\nu,\rho\sigma}(\omega,\vec{q};-\omega',-\vec{q}') = -i \int \theta(t) [T^{\mu\nu}(t,\vec{x}),T^{\rho\sigma}(t',\vec{x}')] e^{i \omega t}e^{-i \vec{q} \cdot\vec{x}} e^{i \omega' t'}e^{-i \vec{q}' \cdot\vec{x}'} dt' dt d^3x d^3x'  \,.
\end{equation}
In Fourier space \eqref{E:deltaT} reads
\begin{align}
\begin{split}
	\delta T^{\mu\nu}(\omega,\vec{q}) &= -\frac{1}{2} \int G_R^{\mu\nu,\rho\sigma}(\omega,\vec{q};-\omega',-\vec{q}') \delta(\vec{q}) h_{lm}(\omega') d\omega'd^3q' + \mathcal{O}(h^2) \\
	&=-\frac{1}{2} \int G_R^{\mu\nu,\rho\sigma}(\omega,\vec{q};-\omega',0)h_{\rho\sigma}(\omega') d\omega' + \mathcal{O}(h^2)\,.
\end{split}
\end{align}
If the Green function is invariant with respect to time translations then we can define
\begin{equation}
	G_R^{\mu\nu,\rho\sigma}(\omega,\vec{q};-\omega',0)=G_R^{\mu\nu,\rho\sigma}(\omega,\vec{q}) \delta(\omega-\omega')
\end{equation}
from which
\begin{equation}
\label{E:deltaTmn2}
	\delta T^{\mu\nu}(\omega,\vec{q}) = -\frac{1}{2} G_R^{\mu\nu,\rho\sigma}(\omega,\vec{q})h_{\rho\sigma}(\omega)+ \mathcal{O}(h^2) 
\end{equation}
follows.
Had the Green function been invariant under spatial translations then we'd use the canonical relation
\begin{equation}
	G_R^{\mu\nu,\rho\sigma}(\omega,\vec{q})  = G_R^{\mu\nu,\rho\sigma}(\omega)\delta(\vec{q}) \,.
\end{equation}

Turning on a time dependant source term for the charge current, $\alpha_\mu$, we find that 
\begin{align}
\begin{split}
\label{E:alldelta}
	\delta T^{\mu\nu}(\omega,\vec{q})  &= -\frac{1}{2} G_R^{\mu\nu,\rho\sigma}(\omega,\vec{q})h_{\rho\sigma}(\omega) - G_R^{\mu\nu,\rho}(\omega,\vec{q})\alpha_{\rho}(\omega) +\mathcal{O}(h^2 ,\alpha h,\alpha^2) \\
	\delta J^{\mu}(\omega,\vec{q}) & = - G_R^{\mu,\nu}(\omega,\vec{q}) \alpha_{\nu}(\omega) - \frac{1}{2} G_R^{\mu,\nu\rho}(\omega,\vec{q})h_{\nu\rho}(\omega) +\mathcal{O}(h^2 ,\alpha h,\alpha^2)
\end{split}
\end{align}
where 
\begin{align}
\begin{split}
	G_R^{\mu,\nu}(\omega,\vec{q})\delta(\omega-\omega') &= - i \int  \theta (t) [J^{\mu}(t,\vec{x}),J^{\nu}(t',\vec{x}')]   e^{i \omega t}e^{-i \vec{q} \cdot\vec{x}} e^{i \omega' t'} dt' dt d^3x d^3x'   \\
	G_R^{\mu,\nu\rho}(\omega,\vec{q}) \delta(\omega-\omega') &= - i \int  \theta (t) [J^{\mu}(t,\vec{x}),T^{\nu\rho}(t',\vec{x}')]    e^{i \omega t}e^{-i \vec{q} \cdot\vec{x}} e^{i \omega' t'} dt' dt d^3x d^3x' \\
	G_R^{\mu\nu,\rho}(\omega,\vec{q}) \delta(\omega-\omega') &= - i \int  \theta (t) [T^{\mu\nu}(t,\vec{x}),J^{\rho}(t',\vec{x}')]   e^{i \omega t}e^{-i \vec{q} \cdot\vec{x}} e^{i \omega' t'} dt' dt d^3x d^3x' \,.
\end{split}
\end{align}
Note that the absence of translation invariance does not allow us to relate $G_R^{\mu,\nu\rho}$ to $G_R^{\nu\rho,\mu}$.

To compute the various Green functions we will use the results of \cite{Sahoo:2010sp} to map fluctuations of the metric and gauge field due to a time dependent source $h_{\mu\nu}e^{-i\omega t}$ and $\alpha_{\mu}e^{-i\omega t}$ to fluctuations of the energy momentum tensor and current on the boundary theory $\delta T^{\mu\nu}$ and $\delta J^{\mu}$. We then use the relations \eqref{E:alldelta} to compute the associated Green functions which eventually lead to \eqref{E:etasfinal1}, \eqref{E:etasfinal2}, \eqref{E:sigmafinal}, and \eqref{E:mixedfinal}. We start with the simplest case of the current response to perturbations of the gauge field and the proceed to the more difficult computation of the stress-stress correlation functions. Many of the details of the computation have been relegated to appendix \ref{A:Correlators}.

\subsection{Current-current correlation functions}
\label{S:Currentcurrent}
To compute current-current correlation functions we add to our background gauge field a small perturbation $\delta A_M = \alpha_M(x,u)e^{-i\omega t}$ such that
\begin{equation}
\label{E:Apertubation}
	A = a dt + b \omega_y + \gamma^{-1}   \alpha_{M}  e^{-i \omega t}  dx^{M} + \mathcal{O}(\gamma^{-2})\,.
\end{equation}
After some massaging the linearized equations of motion for $\alpha_M$ take  the form
\begin{subequations}
\label{E:deltaA}
\begin{align}
\begin{split}
	i \hat{\omega} \alpha_t' & = -  (1-u^2) \hat{\partial}_x \alpha_x' - 2 \hat{k} \hat{\omega} \left(\alpha_- - \alpha_+\right) \hat{b}_1 \\
	\alpha_t'' + \frac{\hat{\partial_1}^2 \alpha_t}{4 u (1-u^2)} & = - \frac{i \hat{\omega} \hat{\partial}_x \alpha_x}{4 u (1-u^2)} + 2 i \hat{k} ( (\alpha_--\alpha_+)\hat{b}_1)' - 2 \hat{b}_1' \hat{\partial}_x \left(\alpha_-+\alpha_+\right)  \\
	\left((1-u^2) \alpha_x'\right)' &-  \frac{\hat{\omega}^2}{4 u (1-u^2)} \alpha_x  = \frac{i \hat{\omega}}{4 u (1-u^2)} \hat{\partial}_x \alpha_t+ 2 i \hat{\omega} \hat{b}_1'  \left(\alpha_- + \alpha_+\right)  \\ 
\end{split}
\end{align}
and
\begin{multline}
	\left((1-u^2) \alpha_\pm'\right)' +  \frac{\hat{\omega}^2}{4 u (1-u^2)} \alpha_\pm - \frac{(\hat{k}\mp i\hat{\partial}_x)^2}{4 u} \alpha_\pm - 4 (\hat{k} \mp i \hat{\partial}_x) \hat{a}_1'  \alpha_\pm 
	\\
	= -4   \hat{b}_1'  \hat{\partial}_x\alpha_t - 4 i \hat{\omega} \hat{b}_1'  \alpha_x \pm 4 i \hat{b}_1 \hat{k} \alpha_t'
\end{multline}
\end{subequations}
where we have defined
\begin{equation}
	\alpha_{\pm} e^{\pm i k x_1} = \alpha_3 \pm i \alpha_2,
	\qquad
	\hat{\partial}_x = \frac{1}{r_+} \frac{\partial}{\partial x}
	\qquad
	\hat{\omega} = \frac{\omega}{r_+}
\end{equation}
and picked the gauge $\alpha_u=0$. In what follows we will carry out a Fourier transform of $\alpha_{\pm}$, $\alpha_x$ and $\alpha_t$ in the $x^1$ direction such that $\hat{\partial}_x = i \hat{q} = i q/r_+$. The first equation in \eqref{E:deltaA} is a constraint equation. Note the similarity between the equation of motion for $\alpha_{\pm}$ and the linearized version of \eqref{E:abequations}. 

The boundary conditions we impose on our fields are that they are ingoing at the horizon and that $\alpha_{i}$ are constant at the asymptotically AdS boundary. In Fourier space this amounts to:
\begin{equation}
	\alpha_x \xrightarrow[u\to 0]{} e_x \delta(q)\,,
	\qquad
	\alpha_{\pm} \xrightarrow[u\to 0]{} e_{\pm} \delta(q\pm k)\,.
\end{equation}
Apart from these six boundary conditions we have a pure gauge solution of the form $\alpha_t = \omega e_t$, $\alpha_x = - q e_t$ with $e_t$ a constant.

It is difficult to solve \eqref{E:deltaA} explicitly. In what follows we will solve them perturbatively in $\hat{\omega}$. Using the standard method of matching the small frequency limit of a near horizon expansion with the near horizon limit of a small frequency expansion, we find that 
\begin{align}
\begin{split}
\label{E:deltaAsol}
	\alpha_{\pm} =& -e_x \delta(q) \hat{b}_1(u) \left( 4  + 2  \frac{i \hat{\omega}}{\hat{b}_1(1)^2} \left(\int_0^1 \frac{\hat{b}_1^2}{t(1+t)}dt + 16 \int_0^1 \frac{\hat{b}_1^4 - \hat{b}_1^2(1)\hat{b}_1^2}{1-t^2} dt +  \arctanh(u) \right) \right) \\
	&+ e_{\pm} \delta(k\pm q) \left(1 + \frac{1}{2}i\hat{\omega} \arctanh(u) + 4 \hat{k} \hat{\omega} \int_u^1 \frac{\hat{b}_1 \beta_1 - \hat{b}_1(1) \beta_1(1)}{1-t^2}dt \right) \\
	&+e_{\mp} \delta(k \mp q) i \hat{\omega} \beta_2 
	+\mathcal{O}(\hat{\omega}^2) 
	 \\
	\alpha_{x} =& e_{x} \delta(q) \left( 1+ \frac{ i \hat{\omega}}{2} \left(\arctanh(u) + 16 \int_u^1 \frac{\hat{b}_1^2 - \hat{b}_{1}(1)^2}{1-t^2} dt \right)\right)  \\
	&+ 2 i  \hat{\omega} (e_- \delta(q-k) + e_+ \delta (q+k)) \left(\hat{b}_1 \arctanh(u)+ \int_u^1 \frac{\hat{b}_1^2 - \hat{b}_{1}(1)^2}{1-t^2} dt \right) +\mathcal{O}(\hat{\omega}^2) \,.
\end{split}
\end{align}
In \eqref{E:deltaAsol} the functions $\beta_i$ are defined via the differential equations 
\begin{subequations}
\label{E:betaieqn}
\begin{multline}
	-u(1-u^2)\beta_1'' + \frac{1}{4}\left(1+32 u \hat{b}_1^2\right)\hat{k}^2 \beta_1 + 2 i u (1-u^2) \hat{b}_1  \hat{k} \beta_2' + 4 i u (1-u^2) \hat{k} \hat{b}_1' \beta_2 \\
	 = i u \hat{b}_1 \hat{k} \left(1 +32 \hat{b}_1^2(1)\right)  
\end{multline}
and
\begin{equation}
	((1-u^2)\beta_2')' + \hat{k} \beta_2 \left( \left(-u^{-1} + 16 \int_0^1 \hat{b}_1^2 dt - 16 \hat{b}_1^2\right) \hat{k} + 8 \hat{\mu} \right) +  4 i \hat{k} \left(\hat{b}_1 \beta_1' - \hat{b}_1' \beta_1  \right) =0
\end{equation}
\end{subequations}
with the boundary conditions $\beta_i(0) = 0$, $\beta_1(1) = \frac{4 i \hat{b}_1}{\hat{k}}$ and $\beta_2(1)$ is finite. The details of the computation leading to \eqref{E:deltaAsol} can be found in appendix \ref{A:JJexpansion}. We have not managed to compute the $\beta_i$ in closed form. In section \ref{S:Perturbative} we will make some progress in this direction by solving \eqref{E:betaieqn} perturbatively in $\Delta$.

Going back to real space and using the prescription of \cite{Sahoo:2010sp} the (consistent) current-current correlator is given by
\begin{equation}
	\frac{i}{\omega} G_{i,j} = 
	 -\frac{N^2 \pi^2 T}{V_5 \gamma} 
		\begin{pmatrix} 
		\frac{1}{8} + s_0  	&  0  &  0  \\
		\left( \frac{i \pi T \Delta}{\omega} - s_1 \right) \sin (k x) & \frac{1}{8} + 2 s_0 - s_2 \cos(2 k x) & s_2 \sin (2 k x)  \\
		\left( \frac{i \pi T \Delta}{\omega} - s_1 \right) \cos (k x) & s_2 \sin(2 k x) & \frac{1}{8}  + 2 s_0 + s_2 \cos(2 k x)
		\end{pmatrix}
\end{equation}
where
\begin{align}
\begin{split}
\label{E:si}
	s_0 & = 2 \hat{b}_1^2(1) \qquad
	s_2  = \frac{1}{4} \beta_2'(0) 
	\\
	s_1 & = \frac{\Delta}{2 \hat{b}_1(1)^2} \left( \int_0^1 \frac{\hat{b}_1^2(t)}{t(1+t)}dt - 16 \int_0^1 \frac{\left(\hat{b}_1(t)^2 - \hat{b}_1^2(1)\right)^2}{1-t^2} dt \right) \,.
\end{split}
\end{align}

\subsection{Stress tensor-current correlation functions}
To compute the stress tensor-current correlator we need to compute the response of the metric to the gauge-field perturbation \eqref{E:Apertubation}. Let us denote the linearized response of the  metric perturbations to the gauge field by $h_{MN}$, i.e., $ds^2 = g_{MN}dx^M dx^N +  h_{MN}dx^M dx^N + \mathcal{O}(\gamma^{-4})$ with $g_{MN}$ given by the line element \eqref{E:helicalansatz}. In what follows we focus on the response of the $xy$ component of the stress tensor. As it turns out the equation of motion for $h_{xy}$ is coupled to that of $h_{ty}$. Working perturbatively in $\omega$ these equations of motion decouple and may be solved. We find, in Fourier space,
\begin{align}
	\frac{\gamma^2 u h_{xy}}{r_+ \omega} = & \frac{1}{4} (e_-+e_+)\delta(q) \left(2 \int_0^u \frac{s \hat{b}_1^2 - \hat{b}_1^2(1)}{1-s^2} ds + \hat{b}_1^2(1)\left(2 \arctanh(u) + \ln(1-u^2)\right) \right) \\
		& + e_x \left(\delta(q-k) + \delta(q+k)\right) \Bigg( \int_0^u \frac{s \hat{b}_1^2 - \hat{b}_1^2(1)}{1-s^2} ds \left(-\frac{1}{4} + 4 \int_0^u \hat{b}_1^2 ds - 4 \hat{b}_1^2(1) + \frac{2 \hat{\mu}}{\hat{k}} \right) \\
		&+ \arctanh(u) \left( - \hat{b}_1 \left(\frac{1}{4}-4 \int_0^1 \hat{b}_1^2 + 4 \hat{b}_1(1)^2 \right) + \frac{1}{\hat{k}} \left( 2 \hat{b}_1 \hat{\mu} + \theta'(1) \right) \right) \\
		&+ \frac{1}{\hat{k}} \int_0^u \frac{\theta' - \theta'(1)}{1-s^2} ds 
		\Bigg) \\
		&- \frac{1}{2} (e_- \delta(q-2k) + e_+ \delta(q+2k) ) \left(\arctanh(u) \hat{b}_1^2 + \int_0^u \frac{s \hat{b}_1^2 - \hat{b}_1^2(1)}{1-s^2} ds \right).
\end{align}
where 
\begin{multline}
	\theta = \mathfrak{h}_2 \int_0^u \mathfrak{h}_1(s) \left( 2 \hat{k}  \left(\int_0^1 \hat{b}_1^2(t) dt - \hat{b}_1(s)^2\right) + \hat{\mu} \right) \hat{b}_1'(s) ds \\
	+ \mathfrak{h}_1 \int_u^1 \mathfrak{h}_2(s) \left( 2 \hat{k}  \left(\int_0^1 \hat{b}_1^2(t) dt - \hat{b}_1^2(s)\right) + \hat{\mu} \right) \hat{b}_1'(s) ds\,,
\end{multline}
and the $\mathfrak{h}_i$ are solutions to
\begin{equation}
\label{E:heqn}
	\left(\frac{\mathfrak{h}_i'}{u} \right)' - \frac{\mathfrak{h}_i \hat{k}^2}{4 u^2 (1-u^2)} = 0
\end{equation}
such that $\mathfrak{h}_1(0)=0$, $\mathfrak{h}_2(0)=1$ and $\mathfrak{h}_2(1)=0$ whereas $\mathfrak{h}_1(1)$ does not vanish. 
	
Once again using the prescription of \cite{Sahoo:2010sp} we find that
\begin{equation}
		G_R^{xy,i}  = \frac{N^2 \pi^4 T^3 i \omega}{2 V_5 \gamma^2} \left(\frac{T}{k} \cos(k x) u_0,\,0,\,u_1\right) + \mathcal{O}(\omega^2 ,\gamma^{-4}) \\
\end{equation}
with
\begin{equation}
\label{E:ui1}
	u_0 = 2 \pi  \int_0^1\left(  2\hat{k}  \left(\int_0^1 \hat{b}_1^2 dt - \hat{b}_1 \right) + \hat{\mu} \right)  \mathfrak{h}_2 \hat{b}_1' ds \qquad
	u_1 =  \frac{1}{2} \hat{b}_1^2(1).
\end{equation}

\subsection{Current-stress tensor correlation functions}
In order to compute the current-stress tensor correlation functions we need to compute the reaction of the gauge field to a metric perturbation. Since we will be perturbing the background metric at order $\gamma^0$ we use the notation,
\begin{equation}
	h_{\mu\nu} = h_{\mu\nu}^{(0)} + \gamma^{-2} h_{\mu\nu}^{(2)}  + \mathcal{O}(\gamma^{-2})\,.
\end{equation}
The Fourier transform of the equations of motion for the order $\gamma^0$ metric perturbations of $h_{ij}$, $i\neq j$ is given by
\begin{equation}
	\left(u(1-u^2)h_{ij}^{(0)\prime}\right)^\prime+\left( \frac{1}{4}\frac{\hat{\omega}}{1-u^2}- \frac{(1+u^2)}{u}\right)h_{ij}^{(0)} = 0\,,\qquad (i\neq j)\\
\end{equation}
with boundary conditions
\begin{equation}
	h_{xy}^{(0)}(u) =  h_0 \delta(q) \frac{1}{u} e^{i q x} + \mathcal{O}(u^0)
	\qquad
	h_{yz}^{(0)}(u) = h_1 \delta(q) \frac{1}{u} e^{i q x} + \mathcal{O}(u^0)
\end{equation}
and that $h_{xy}$ and $h_{yz}$ are incoming at the horizon. (Note that $\lim_{r  \to \infty} h_{xy}(r)/r^2 = h_0/r_+^2$.) A standard computation gives us
\begin{equation}
	h_{xy}^{(0)} = h_0 \delta(q) \left(\frac{1}{u} + \frac{1}{4}i \hat{\omega} u \right) + \mathcal{O}(u^2, \hat{\omega})
	\qquad
	h_{yz}^{(0)} = h_1 \delta(q) \left(\frac{1}{u} + \frac{1}{4}i \hat{\omega} u \right) + \mathcal{O}(u^2, \hat{\omega})\,.
\end{equation}

Using the same notation as in \ref{S:Currentcurrent}, the order $\gamma^{-1}$ equations of motion for the response of the gauge field to the metric perturbation take the form \eqref{E:deltaA} but with extra sources for the $\alpha_\mu$. 
As described in appendix \ref{A:stress-stress} we find that the series expansion of the solution, in real space, is given by
\begin{align}
\begin{split}
\label{E:afromh}
	\alpha_{\pm} =& \hat{h}_0 i \hat{\omega} u \left( e^{\pm i k x} \beta_4'(0) - e^{i k x} \hat{b}_1^2(1) \right) \\
	& + \hat{h}_1 u \left( \frac{\Delta}{2}  \delta(q\pm 2k)+  i \hat{\omega}  \left(  \beta_5'(0) \delta(q \pm 2k)  + \beta_6'(0) \delta(q \mp 2k)  \right) \right)  
	+ \mathcal{O}(u^2,\hat{\omega})  \\
	\alpha_{x} = &  \hat{h}_0 \cos(k x)  \Delta u   + \mathcal{O}(u^2,\hat{\omega^2})
\end{split}
\end{align}
where similar to $\beta_1$ and $\beta_2$, the $\beta_i$'s with $i=4,5,6$ are solutions to a set of linearly coupled equations whose explicit form is given in \eqref{E:betaieqnV2}. 
	
Using the prescription of \cite{Sahoo:2010sp} we find that the current-stress tensor correlators are given by:\footnote{Note that $\frac{1}{\sqrt{g}} \frac{\delta^2 W}{\delta g_{\mu\nu}(x) \delta A_i(y)}$ and $\frac{1}{\sqrt{g}} \frac{\delta^2 W}{\delta g_{\mu\nu}(x) \delta A^i(y)}$ differ by contact terms which will, in the current context, contribute to frequency independent terms in the Greens function.} 
\begin{subequations}
\label{E:mixedcorrelators}
\begin{equation}
	G_R^{i,xy} = \frac{N^2 \pi^4 T^3}{4 V_5 \gamma}  \left( \Delta \cos(k x),\,\frac{i\omega}{T} \sin(2 k x)u_2 ,\, -\frac{i \omega}{T} u_3 + \frac{i\omega}{T} \cos(2 k x) u_2 \right) + \mathcal{O}(\omega^2,\gamma^{-2}) \,,
\end{equation}
and
\begin{equation}
		G_R^{i,yz} = \frac{N^2 \pi^4 T^3}{4 V_5 \gamma}  \left( 0,\,-\sin( k x) g + i \frac{\omega}{T} \sin(3 k x) u_5,\, \cos( k x)g + i \frac{\omega}{T} \cos(3 k x) u_5 \right)  +  \mathcal{O}(\omega^2 ,\gamma^{-4}) \,,
\end{equation}
\end{subequations}
where
\begin{equation}
	g = -\frac{\Delta}{2} + i \frac{\omega}{T} u_4
\end{equation}
and
\begin{equation}
\label{E:ui2}
	u_2 = \frac{\beta_4'(0)}{\pi}
	\qquad
	u_3 = \frac{\hat{b}_{1}(1)^2}{\pi}
	\qquad
	u_4 = \frac{\beta_5'(0)}{\pi}
	\qquad
	u_5 = \frac{\beta_6'(0)}{\pi}\,.
\end{equation}	
	
\subsection{Stress tensor-stress tensor correlation functions}

Using the results of the previous section we can now work to order $\gamma^{-2}$ and compute the response of the metric to small perturbations including mediation via the gauge field. Using the method of matched asymptotic expansions as described in appendix \ref{A:stress-stress} we find that
\begin{align}
\begin{split}
	h_{xy}^{(2)} =& h_0 \left(\delta (q+2 k)+ \delta(q-2k) \right) \frac{v_2}{u} 
	- i \hat{\omega} h_0  \frac{1}{4} \left(\delta( 2k+q) + \delta(2k-q) \right) u^{-1} \int_0^u \ln(1-t^2) v_2'  dt \\
	&+ i \hat{\omega} h_0 \delta(q)u^{-1} \left(F_1(u) + \int_0^u \frac{s \int_s^1 F_2(t) dt }{1-s^2} ds \right)  \\
	h_{yz}^{(2)}  =& h_1 \delta(q)  u^{-1} \left(F_3 + \int_0^u \frac{s \int_s^1 F_4 dt}{1-s^2}ds \right) \\
	&- \frac{i \hat{\omega} h_1}{16 u}  \left(\delta(q+4k)+\delta(q-4k)\right) \left(\mathfrak{c}_1 \int_u^1 F_5 \mathfrak{c}_2 ds + \mathfrak{c}_2 \int_0^u F_5 \mathfrak{c}_1 ds \right)  
\end{split}
\label{E:hhGamma2SOl}
\end{align}
where the $F_i(u)$ are given by explicit integrals of $\hat{b}_1$ in equations \eqref{E:F1}, \eqref{E:F2}, \eqref{E:F3} and \eqref{E:F4} of appendix \ref{A:stress-stress}.
The $\mathfrak{c}_i$ are the solutions to the homogeneous equation
\begin{equation}
\label{E:cequation}
	\left(	\left( \frac{1}{u} - u \right)  \left( u \mathfrak{c}_i \right)^\prime \right)' - \frac{4 \hat{k}^2}{u} \mathfrak{c}_i = 0 \,.
\end{equation}
where $\mathfrak{c}_1 = u^2 + \mathcal{O}(u^3)$, $\mathfrak{c}_2 = 1 + \mathcal{O}(u)$ and $\mathfrak{c}_2$ is finite at the horizon while $\mathfrak{c}_1$ diverges there and
\begin{equation}
\label{E:Qval}
	F_5 = \frac{3 \hat{k}^2}{u} \hat{b}_1 \beta_6 - 4 (1-u^2) \hat{b}_1' \beta_6' \,.
\end{equation}

Using the prescription of \cite{Sahoo:2010sp} we obtain
\begin{equation}
	\lim_{\omega \to 0} \frac{\hbox{Im}(G_R^{xy,xy})/\omega}{s} = \frac{1}{4\pi} + \frac{t_0}{2\pi \gamma^2} + \mathcal{O}(\gamma^{-3})
\end{equation}
where
\begin{equation}
	t_0	= -\frac{1}{2}\hat{h}_2(1) + \int_0^1 F_2 dt + F_1''(0) \,.
\end{equation}
and
\begin{equation}
	\lim_{\omega \to 0} \frac{\hbox{Im}(G_R^{yz,yz})/\omega}{s} = \frac{1}{4\pi} + \frac{t_1+t_2 \cos(4 k x)}{2\pi \gamma^2} + \mathcal{O}(\gamma^{-3})
\end{equation}
where
\begin{equation}
\label{E:ti}
	t_1	= -\frac{1}{2}\hat{h}_2(1) + \int_0^1 F_4 dt + F_3''(0) \qquad
	t_2   = -\frac{1}{4} \int_0^1 F_5 \mathfrak{c}_2 ds \,.
\end{equation}

\subsection{Perturbative solution near $\Delta=0$.}
\label{S:Perturbative}
Using the perturbative expansion described in section \ref{S:PerturbativeBgd} many of the correlation functions described in this section may be computed explicitly. Indeed let us consider a compactified $x$ direction and fixed chemical potential so that \eqref{E:lowest} describes the helical phase close to the phase transition  (denoted by the red dot in figure \ref{F:critcurve}). Inserting 
\eqref{E:abexpansion} and \eqref{E:kmuexpansion} into the expressions for $s_i$, $t_i$ and $u_i$, (equations \eqref{E:si}, \eqref{E:ti}, \eqref{E:ui1} and \eqref{E:ui2}) we find that
\begin{align}
\begin{split}
	s_0 &= 2 (1+\sqrt{2})^2 \Delta^2 + \mathcal{O}(\Delta^3)
	\\
	s_1 &= \left(6 - \frac{25}{3\sqrt{2}} + \sqrt{2} \ln(64) - \ln(256 \sqrt{2}) \right) \Delta + \mathcal{O}(\Delta^2) 
	\\
	t_0 &= \frac{1}{900} \left(-427-225 \sqrt{2}+210 \log (2)\right)\Delta^2 + \mathcal{O}(\Delta^3)  \sim -0.66 \Delta^2 
	\\
	t_1 & = \frac{1}{8} \left(3+2 \sqrt{2}\right)\Delta^2 + \mathcal{O}(\Delta^3) \sim 0.72 \Delta^2
	\\
	u_1 & = \frac{1}{2} (1+\sqrt{2})^2 \Delta^2 + \mathcal{O}(\Delta^3)
	\\
	u_3 & = \frac{1}{\pi} (1+\sqrt{2})^2 \Delta^2 + \mathcal{O}(\Delta^3) 
	\\
	u_4 & = \frac{1}{48\pi} \Delta + \mathcal{O}(\Delta^3)\,.
\end{split}
\end{align}

The remaining terms $s_2$, $u_0$, $u_2$, $u_5$, $t_2$ can only be computed numerically. It is relatively straightforward to obtain a numerical solution for $\mathfrak{h}_i$ defined in \eqref{E:heqn} and then to evaluate $u_0$ using \eqref{E:ui1}. We find
\begin{equation}
	u_0 \sim 0.14 \Delta + \mathcal{O}(\Delta^2) \,.
\end{equation}

In order to compute $s_2$, $u_2$, $u_4$ and $u_5$ we solve the equations for the $\beta_i$'s defined in \eqref{E:betaieqn} and \eqref{E:betaieqnV2} perturbatively in $\Delta$. Let
\begin{equation}
\label{E:betaiexpansion}
	\beta_i = \sum_{n=0} \beta_i^{(n)} \Delta^n\,.
\end{equation}
Inserting \eqref{E:betaiexpansion} into \eqref{E:betaieqn} and \eqref{E:betaieqnV2} we find that the equations for the $\beta_i^{(n)}$'s decouple and becomes a set of inhomogenous linear equations whose detailed form can be found in appendix \ref{A:Numerical}. Inserting the numerical values of the $\beta_i$'s into \eqref{E:si} and \eqref{E:ui2} we find  
\begin{align}
\begin{split}
\label{E:sunumerical}
	s_2 & \sim - 10^{-2} \Delta^2 + \mathcal{O}(\Delta^3) \\
	u_2 & \sim 0.07 \Delta^2 + \mathcal{O}(\Delta^3) \\
	u_5 & \sim 4 \times 10^{-4} \Delta^3 + \mathcal{O}(\Delta^4)\,.
\end{split}
\end{align}
Evaluating $\mathfrak{c}_2$ numerically and using \eqref{E:ti}, we find
\begin{equation}
\label{E:t2numerical}
	t_2 \sim 4.4 \times 10^{-5} \Delta^4 + \mathcal{O}(\Delta^5)\,.
\end{equation}

\section{Discussion}
\label{S:Discussion}
In this work we have computed various low frequency correlation functions in a thermally equilibrated but non isotropic spatially modulated background. These correlators should respect the $E(2)\times \mathbb{Z}_2$ symmetry of the background. In particular, if we denote the generators of the $E(2)$ symmetry by $\partial_{y}$, $\partial_{z}$ and $\xi = \partial_x - k \left(z \partial_y - y \partial_z\right)$ then all correlators should have vanishing Lie derivative with respect to these three generators. 

Consider the conductivity tensor $\sigma_{ij}$. Requiring that $\sigma$ is invariant under infinitesimal translations in the $y$, $z$ and $\xi$ directions and also under $x\to -x$ and $z \to -z$ implies that
{\small
\begin{equation}
\label{E:Cgeneral}
	\sigma = \begin{pmatrix}
		\sigma_0 & \sigma_1 \sin(k x) & \sigma_1 \cos(k x)   \\
		 \sigma_2 \sin(k x) & \sigma_3 - \sigma_4 \cos(2 k x)   & \sigma_4 \sin(2 k x) \\
		\sigma_2 \cos(k x) &  \sigma_4 \sin(2 k x) & \sigma_3 + \sigma_4 \cos(2 k x) 
		\end{pmatrix}\,.
\end{equation}
}
where the $\sigma_i$ can also be functions of $k$. In terms of the helical one forms $\omega_y$ and $\omega_z$ defined in \eqref{E:omegayz} equation \eqref{E:Cgeneral} takes the form
\begin{equation}
\sigma = \sigma_0 \, dx \otimes dx + \sigma_1 \, dx \otimes d\omega_z + \sigma_2 \, \omega_z \otimes dx + (\sigma_3 - \sigma_4) \, \omega_y \otimes \omega_y + (\sigma_3 + \sigma_4) \, \omega_z \otimes \omega_z\,.
\end{equation}
The general form of the conductivity matrix \eqref{E:Cgeneral} agrees with our explicit computation \eqref{E:sigmafinal} upon setting
\begin{equation}
	\sigma_1  =0
\end{equation}
and
\begin{align}
\begin{split}
	\sigma_0 &= -\frac{N^2 \pi^2 T}{V_5 \gamma} \left( \frac{1}{8} + 2 \hat{b}_1^2(1)  \right) \\
	\sigma_2 &= -\frac{N^2 \pi^2 T \Delta}{V_5 \gamma} \left(\frac{i \pi T }{\omega}- \frac{1}{2 \hat{b}_1(1)^2} \left( \int_0^1 \frac{\hat{b}_1^2}{t(1+t)}dt + 16 \int_0^1 \frac{\hat{b}_1^4 - \hat{b}_1^2(1)\hat{b}_1^2}{1-t^2} dt \right)\right) \\
	\sigma_3 &= -\frac{N^2 \pi^2 T}{V_5 \gamma}\left(\frac{1}{8} + 4 \hat{b}_1^2(1)\right) \\
	\sigma_4 &= -\frac{N^2 \pi^2 T}{4 V_5 \gamma} \beta_2'(0) \,.
\end{split}
\end{align}

Setting $\Delta=0$ we recover the isotropic conductivity matrix associated with a thermal state in the probe limit (compare with, e.g., the results in section 4 of \cite{Son:2006em}). Once $\mathcal{O}(\Delta)$ corrections are taken into account, the conductivity matrix acquires a non trivial spatially modulated structure and, in addition, a divergent $1/\omega$ contribution to $\sigma_2$. Using standard arguments (see, e.g., \cite{Horowitz:2012ky}) the latter pole can be associated with a delta function contribution to the DC conductivity due to translation invariance in the $y$ and $z$ directions.\footnote{The careful reader will note that at order $\mathcal{O}(\Delta^0)$ the helical phase is absent and translation symmetry exists in all three spatial directions. The reason one does not notice a pole structure in $\sigma_0$ and $\sigma_3$ in the $\Delta \to 0$ limit is that this pole has strength $\mu^2$ and is therefore not observable in the probe limit.}

Requiring that the stress-stress two point function also respect the $E(2)\times \mathbb{Z}_2$ symmetry results in
\begin{align}
\begin{split}
\label{E:etageneral}
	G_R^{xy,xy} &=  \tau_0 + \tau_1 \cos(2 k x)   \\
	G_R^{yz,yz} & = \tau_2 +  \tau_3 \cos(4 k x)\,.
\end{split}
\end{align}	
Comparing \eqref{E:etageneral} with \eqref{E:etasfinal1} and \eqref{E:etasfinal2} and noting that the entropy of the helical configuration is not modulated we observe that
\begin{equation}
	\tau_1 =  0\,.
\end{equation}
The non vanishing of $\tau_4$ implies that the zero frequency limit of $\hbox{Im}(G_R^{yz,yz})/(\omega s)$ is spatially modulated. In addition the non zero $\mathcal{O}(\Delta)$ contributions to $\tau_0$ and $\tau_2$ modify the canonical relation \eqref{E:etas}. 

The vanishing of $\tau_1$ in \eqref{E:etageneral} implies that $G_R^{xy,xy}$ is not modulated. This may be contrasted with the modulation of the bulk metric component $g_{xy}$ in response to a boundary metric perturbation in the $xy$ direction. The reason that the modulation of $g_{xy}$ does not contribute to the $G_R^{xy,xy}$ correlation function is that the perturbed, modulated, contribution to the bulk metric dies off too quickly near the boundary. A similar effect leads to the vanishing of $\sigma_1$ in \eqref{E:Cgeneral}.

As noted in the introduction, inhomogenous backgrounds will generically  violate \eqref{E:etas}, but one should also take into account that in such backgrounds the standard hydrodynamical relation between shear viscosity and the stress-stress two point function is  inaccurate. In the hydrodynamic limit one considers long wavelength perturbations around an equilibrated configuration. If the equilibrated configuration is isotropic then the tensor structure of the response of the stress tensor to gradients in the fluid velocity is tightly constrained. In a non isotropic background this tensor structure is not only less constrained but one should also consider the response of the stress tensor to long wavelength perturbations of the symmetry breaking vector (see, e.g., \cite{Jain:2014vka} for a recent explicit analysis). In the helical configuration we are considering a long wavelength variation of, say, the temperature will, according to figure \ref{F:critcurve}, likely induce a spatial variation of $k$. Thus, in the hydrodynamical description of the fluid the stress tensor may respond to variations of $k$. Such a response may, or may not, affect the relation between stress-stress two point functions and shear viscosity. In this work we have referred to the response of the stress tensor to a spatial metric perturbation as a generalized ``shear''.

We have also mentioned in the introduction that there are a handful of other instances where the stress-stress two point function violates the bound \eqref{E:etas} \cite{Erdmenger:2010xm,Basu:2011tt,Erdmenger:2011tj,Rebhan:2011vd}. As is the case in this work, most violations of \eqref{E:etas} in the context of two-derivative gravity have been exhibited in theories whose boundary dual is ill defined or unknown at best (\cite{Rebhan:2011vd} being an exception). In the present context we point out that the consistent truncation of type  IIB supergravity backgrounds describing spinning D3 branes on the tip of a Calabi-Yau cone will always yield an action of the form \eqref{E:bulkaction} with $\gamma = 1/4\sqrt{3}$ \cite{Gubser:2009qm} (see also \cite{Buchel:2006gb}). More generally, any consistent truncation of 11 dimensional supergravity or type IIB supergravity to AdS${}_5$ will also reduce to a Chern-Simons term with coefficient $\gamma=1/4\sqrt{3}$ \cite{Gauntlett:2006ai,Gauntlett:2007ma}. Thus, the probe limit used in this work should be thought of as a toy model. It would be interesting to carry out the current analysis in a setting whose gravity dual is well defined, such as the Sakai-Sugimoto model \cite{Sakai:2004cn,Aharony:2006da,Bergman:2007wp,Ooguri:2010xs}, and then demonstrate a violation of the bound \eqref{E:etas} in a fully controlled string-theoretical setting.

\section*{Acknowledgments}
We thank S. Cremonini, J. Gauntlett and M. Field for useful discussions. OO and AY are  supported by the ISF under grant number 495/11, by the BSF under grant number 2014350, by the European commission FP7, under IRG 908049 and by the GIF under grant number 1156/2011.

\begin{appendix}

\section{Details regarding the computation of the correlation functions}
\label{A:Correlators}
\subsection{Current-current correlators}
\label{A:JJexpansion}
\begin{figure}[hbt]
\includegraphics[width = \linewidth]{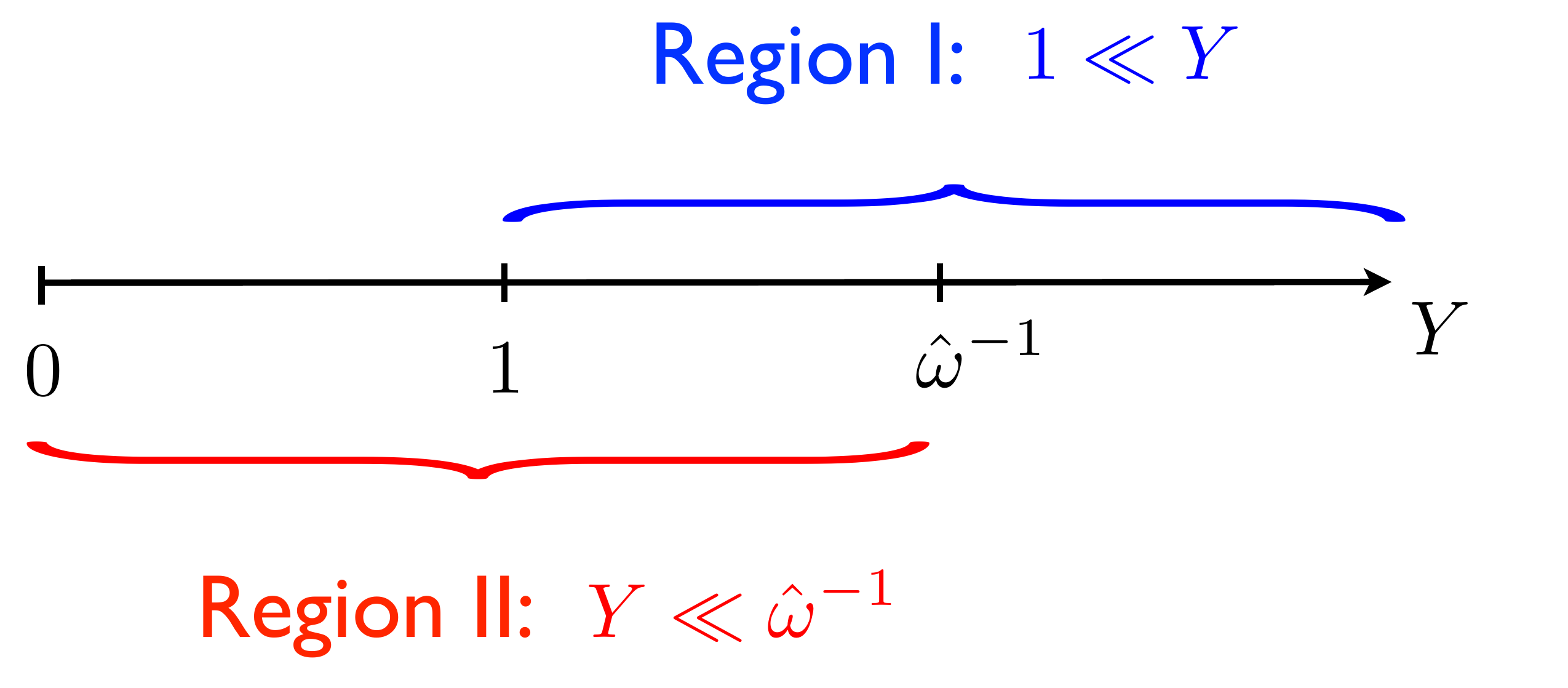}
\caption{\label{F:region12}  A schematic diagram of the two asymptotic regions described in the text. The coordinate $Y$ ranges from $0$ at the asymptotic boundary to $\infty$ at the horizon. Region I is the near horizon region while region II is an expansion valid for values of $Y$ which are parametrically smaller than the frequency.}
\end{figure}

The equations of motion for perturbations of the gauge field in response to an external source are governed by \eqref{E:deltaA}. These equations can be solved in a perturbative expansion in $1/\gamma$ and a matched asymptotic expansion between the near horizon region and a small $\omega$ expansion. We denote
\begin{equation}
	e^{-Y} = (1-u^2)
\end{equation}
so that $Y\to\infty$ is the horizon and $Y=0$ is the boundary of space-time. Let us define the near horizon region (region I) as the region where $Y \gg 1$ and the small frequency region (region II) as the region where $Y \ll \hat{\omega}^{-1}$. These regions have overlap as long as $\hat{\omega} \ll 1$. See figure \ref{F:region12}.

Near the horizon, $Y \to \infty$, the equations of motion for the spatial components of $\alpha$ take the form of a harmonic oscillator \cite{Gubser:2008vz},
\begin{align}
\begin{split}
\label{E:regionI}
	\partial_Y \alpha_t &= \mathcal{O}(e^{-Y}) \\
	\partial_Y^2 \alpha_x + \left(\frac{\hat{\omega}}{4}\right)^2 \alpha_x + \frac{\hat{q}\hat{\omega}}{16} \alpha_t & = \mathcal{O}(e^{-Y}) \\
	\partial_Y^2 \alpha_\pm + \left(\frac{\hat{\omega}}{4}\right)^2 \alpha_\pm & = \mathcal{O}(e^{-Y})\,.
\end{split}
\end{align}
Imposing ingoing boundary conditions at the horizon we find the expansion
\begin{equation}
	\alpha_t^{I} = \frac{\alpha^{\infty}_t\hat{\omega}}{\hat{q}} + \mathcal{O}(e^{-Y})
	\qquad
	\alpha_x^{I} = \alpha^{\infty}_x e^{\frac{i \hat{\omega}}{4}Y} - \alpha_t^{\infty} + \mathcal{O}(e^{-Y})
	\qquad
	\alpha_{\pm}^{I} = \alpha^{\infty}_{\pm} e^{\frac{i \hat{\omega}}{4}Y}  + \mathcal{O}(e^{-Y})\,,
\end{equation}
which is a good approximation to $\alpha_{\mu}$ in region I.
We emphasize that, at this point, the $\alpha^{\infty}_\mu$ may be arbitrary functions of $\hat{\omega}$ and $\hat{q}$. 

In region II we may solve the equations of motion perturbatively in $\hat{\omega}$. Let us denote
\begin{equation}
	\alpha_{\mu}^{II} = \sum_n \alpha^{(n)}_{\mu} (u) (i \hat{\omega})^n\,.
\end{equation}
The equations of motion at order $n$ take the form
\begin{subequations}
\label{E:smallw}
\begin{multline}
	u(1-u^2) \alpha_t^{(n)\prime\prime} - \frac{1}{4} \alpha_t^{(n)} \hat{q}^2 - 2 i u (1-u^2) \hat{k} \left(\hat{b}_1 \left(\alpha_+^{(n)} - \alpha_-^{(n)}\right) \right)' \\
		- 2 i u (1-u^2) \hat{q} \hat{b}_1'  \left(\alpha_+^{(n)} + \alpha_-^{(n)}\right) = s_t^{(n)} 
\end{multline}
\begin{align}
	((1-u^2)\alpha_\pm^{(n)\prime})' - \frac{(\hat{k} \pm \hat{q})(\hat{k} \pm \hat{q} + 16 u \hat{a}_1')}{4u} \alpha_\pm^{(n)}+ 4 i \hat{k} \left(\hat{b}_1 \alpha_t^{(n)}\right)' & = s^{(n)}_\pm \\
	\left((1-u^2) \alpha_x^{(n)\prime}\right)' &= \frac{s_x^{(n)\prime}}{\hat{q}}  \\
	(1-u^2) \hat{q} \alpha_x^{(n)\prime} & = {s}_x^{(n)} 
\end{align}
\end{subequations}
with boundary conditions such that
\begin{equation}
\label{E:BCsIIa}
	\alpha_t^{(0)}(0) = 0
	\qquad
	\alpha_x^{(0)}(0) = e_x \delta(q)
	\qquad
	\alpha_\pm^{(0)}(0) = e_\pm \delta(q \pm k)\,.
\end{equation}
and $\alpha^{(n)}_{\mu}(0) = 0$ for $n>1$. Note that the second order equation for $\alpha_x$ follows from the first order one.

In the region $1 \ll Y \ll \hat{\omega}^{-1}$ both solutions should be approximately valid. Thus, we expect that the small $\hat{\omega}$ expansion of the solution in region I will coincide with the large $Y$ expansion of region II,
\begin{equation}
\label{E:BCsIIb}
	\lim_{u\to 1} \alpha_\mu^{II} = \lim_{\hat{\omega}\to 0} \alpha_{\mu}^{I} \,. 
\end{equation}
Equations \eqref{E:BCsIIb} serve as the second set of boundary conditions on \eqref{E:smallw} and \eqref{E:regionI}. 
	
We will now solve \eqref{E:smallw} perturbatively in $\omega$. At order $\omega^0$ we find that the sources $s_\mu$ vanish. Since the equation of motion for $\alpha_{\pm}$ coincides with \eqref{E:abequations} when $q=0$ we find
\begin{equation}
	\alpha_\pm^{(0)} = e_\pm \delta(q\pm k) + \delta(q) x_0 \hat{b}_1
	\qquad
	\alpha_x^{(0)} = e_x \delta(q)
	\qquad
	\alpha_t^{(0)} = 0\,.
\end{equation}
The overall coefficient $x_0$ will be determined shortly.

At linear order in $\omega$ we find that the solution to the equations of motion takes the form
\begin{align}
\notag
	\alpha^{(1)}_{\pm} =&  \delta(q) (x_1- 2 e_1 \arctanh(u)) \hat{b}_1(u)  + e_{\pm} \delta(k\pm q) \left( \frac{1}{2} \arctanh(u) + 4 i \hat{k} \int_0^u \frac{\hat{b}_1 \beta_1 - \hat{b}_1(1) \beta_1(1)}{1-t^2}dt \right) \\
\notag
	&+e_{\mp} \delta(k \mp q) i \hat{\omega} \beta_2 
	 \\
	\alpha_{x}^{(1)} =& e_{x} \delta(q)\frac{ 1}{2} \left(   \arctanh(u) - 16 \int_0^u \frac{\hat{b}_1^2 - \hat{b}_{1}(1)^2}{1-t^2} dt \right)  \\
\notag
	&+ 2 (e_- \delta(q-k) + e_+ \delta (q+k)) \left(\hat{b}_1(1) \arctanh(u)+ \int_0^u \frac{\hat{b}_1^2 - \hat{b}_{1}(1)^2}{1-t^2} dt \right) 
\end{align}
where $\beta_i$ satisfy \eqref{E:betaieqn} and 
\[
	x_0 = -4 e_x
\]
has been determined from matching the order $\omega$ expansion in region II with the near horizon expansion in region I. A similar analysis involving the $\mathcal{O}(\hat{\omega}^2)$ terms determines $x_1$ leading to \eqref{E:deltaAsol} in the main text.

\subsection{Stress-stress correlators}
\label{A:stress-stress}
We will be interested in the response of the metric to perturbations in its $g_{xy}$ and $g_{yz}$ components. For ease of reference we will treat each of these cases separately.

\subsubsection{Perturbations in $g_{xy}$}
\label{A:stress-stress1}
To compute the response of the metric to a metric perturbation in the $xy$ direction we will use the method of matched asymptotic expansions described in section \ref{A:JJexpansion} . In the large $Y$ limit, the equation of motion for the linear response of the metric to a perturbation in the $xy$ direction, $h_{xy}$, is coupled to that of a perturbation in the $ty$ direction, $h_{ty}$. The homogeneous version of these equations takes the form
\begin{equation}
	\partial_Y^2 h_{xy}'' + \left(\frac{\omega}{4\pi T}\right)^2 h_{xy} = - \frac{\omega q}{(\pi T)^2}h_{ty} + \mathcal{O}(e^{-Y})
	\qquad
	h_{ty}' = \mathcal{O}(e^{-Y})
\end{equation}
with   
\begin{equation}
\label{E:Tdef}
	\pi T = -\frac{f(1)g'(1)}{2r_+}
\end{equation}
the Hawking temperature and also the temperature of the boundary field theory.
The asymptotic (large $Y$) behavior of the ingoing modes is then
\begin{equation}
	h_{ty} = \frac{h_{ty}^{\infty} \omega}{q} + \mathcal{O}(e^{-Y})
	\qquad
	h_{xy} = h_{xy}^{\infty} e^{\frac{i\omega}{4\pi T}Y} - h_{ty}^{\infty} + \mathcal{O}(e^{-Y})\,.
\end{equation}

To solve for the metric perturbations in region II, we first work in the probe limit where
\begin{equation}
\label{E:hprobe}
	h_{\mu\nu} = h_{\mu\nu}^{(0)} + \frac{1}{\gamma^2} h^{(2)}_{\mu\nu} + \mathcal{O}(\gamma^{-4}).
\end{equation}
As in the previous section we will denote a small frequency  expansion of the perturbations, valid in region II, via
\begin{equation}
\label{E:hexpansion}
	h^{(0)}_{\mu\nu} = \sum_{n} h_{\mu\nu}^{(0,n)}\hat{\omega}^n
	\qquad
	h^{(2)}_{\mu\nu} = \sum_{n,m} h_{\mu\nu}^{(2,n)}\hat{\omega}^n\,.
\end{equation}
(Note that the modulated phase is observable only at order $\mathcal{O}(\gamma^{-1})$.) At order $\gamma^0$ the equations of motion for $h^{(0)}_{xy}$ decouple from the rest of the equations and we get
\begin{align}
\begin{split}
	\left( (u^{-1} - u) \left(u h^{(0,n)}_{xy}\right)' \right)'  = \sigma_{xy}^{(n)}\,.
\end{split}
\end{align}
Using $\sigma_{xy}^{(0)} = \sigma_{xy}^{(1)} = 0$ and matching the small frequency solution to the large $Y$ solution we find that
\begin{equation}
\label{E:hxyw0}
	u h_{xy}^{(0,0)} = \delta(q) h_0
	\qquad
	u h_{xy}^{(0,1)} = -\frac{i}{4} h_0 \delta(q) \ln (1-u^2) \,.
\end{equation}
where we have imposed boundary conditions such that $\lim_{u\to0} u h_{xy}(u) =  \delta(q) h_0 $.

At order $\gamma^{-1}$ the metric does not get corrected but the gauge field does get sourced by the metric. Denoting the gauge field perturbations by $\alpha_{\mu}$, i.e., $A = a dt + \omega_y b + \gamma^{-1} \alpha_{\mu} dx^{\mu}$ we obtain, following an analysis identical to the one described in the previous section, 
\begin{equation}
	\alpha_t^{(0)} = \alpha_{\pm}^{(0)} = 0
	\qquad
	\alpha_x^{(0)} = \frac{1}{2 }\hat{h}_0 \hat{b}_{1} \left(\delta (q-k) + \delta (q+k)\right)
\end{equation}
and 
\begin{align}
\begin{split}
	\alpha_t^{(1)} &=  \hat{h}_0 \left(\delta(q+k) - \delta(q-k) \right) \beta_3  \\
	\alpha_x^{(1)} & = -\frac{1}{8} \left(\delta(q+k) + \delta(q-k) \right)\hat{h}_0 \int_0^u \ln(1-t^2) \hat{b}'_1(t) dt \\
	\alpha_{\pm}^{(1)} & = \hat{h}_0 \delta(q \mp k) \beta_4 + \hat{h}_0   \delta(q \pm k)  \left( 4 i \hat{k} \int_0^u \frac{\hat{b}_1(t) \beta_3(t) - \hat{b}_1(1) \beta_3(1)}{1-t^2} dt - \int_0^u \frac{\hat{b}_1^2(t) - \hat{b}_1^2(1)}{1-t^2} dt \right)
\end{split}
\end{align}
with $\beta_3$ and $\beta_4$ satisfying the differential equation
\begin{subequations}
\label{E:betaieqnV2}
\begin{multline}
	-u(1-u^2)\beta_3'' + \frac{1}{4}\left(1+32 u \hat{b}_1^2\right)\hat{k}^2 \beta_3 + 2 i u (1-u^2) \hat{b}_1  \hat{k} \beta_4' + 4 i u (1-u^2) \hat{k} \hat{b}_1' \beta_4 \\
	 = -\frac{1}{8} i  \hat{b}_1 \left(1+16 u \hat{b}_1^2(1) + 16 u \hat{b}_1^2 \right) \hat{k} 
\end{multline}
\begin{equation}
	((1-u^2)\beta_4')' + \hat{k} \left( \left(-u^{-1} + 16 \int_0^1 \hat{b}_1^2(t) dt - 16 \hat{b}_1^2\right) \hat{k} + 8 \hat{\mu} \right) \beta_4 +  4 i \hat{k} \left(\hat{b}_1 \beta_3' - \hat{b}_1' \beta_3  \right) = -  \left(\hat{b}_1^2 \right)' \,.
\end{equation}
\end{subequations}
The boundary conditions we impose on the $\beta_i$ are that they vanish at the boundary, are finite at the horizon and, in particular, that $\beta_3(1) = -{i \hat{b}_1(1)} / {2 \hat{k}}$. The latter condition follows from \eqref{E:BCsIIb}.

At order $\gamma^{-2}$ we find the set of equations
\begin{align}
\begin{split}
\label{E:hxyhtyeqn}
	\left(1-u^2 \right)) \hat{q} (u h_{xy}^{(2,n)})' &=  \sigma_{xy}^{(n)} \\
	\left(	\left( u(1-u^2 ) \right)h_{xy}^{(2,n)\prime}\right)' - \frac{(1+u^2)}{u} h_{xy}^{(2,n)} & = \left( \frac{\sigma_{xy}^{(n)}}{u} \right)' \frac{u}{q} \\
	\left(u h_{ty}^{(2,n)\prime}\right)^{\prime} - \left( \frac{1}{u} + \frac{\hat{q}^2}{4(1-u^2)}\right) h_{ty}^{(2,n)} & = \sigma_{ty}^{(n)} \,.
\end{split}
\end{align}
where the source $\sigma_{xy}^{(n)}$ depends on $h_{ty}^{(2,m)}$ and $h_{xy}^{(2,m)}$ with $m<n$.

At order $\hat{\omega}^0$ the solution to \eqref{E:hxyhtyeqn} is given by
\begin{equation}
	h_{xy}^{(2,0)} = h_0 \left( \delta(q-2k) + \delta(q+2k) \right) \frac{v_2}{u}
	\qquad
	h_{ty}^{(2,0)} = x_0 u\,,
\end{equation}
where $x_0$ is yet to be determined.
At the next order in $\hat{\omega}$ we find
\begin{equation}
 	h_{xy}^{(2,1)} = h_0 u^{-1} \left( \delta(q) \left( F_1 + \int_0^u \frac{s \int_s^1 F_2 dt }{1-s^2} ds\right)  -\frac{1}{4}  \left( \delta(q-2k)+\delta(q+2k) \right) \int_0^u \ln(1-s^2) v_2' ds \right)
\end{equation}
solves the equation of motion with
\begin{align}
\begin{split}
\label{E:F1}
	F_1 = & \int_0^u \frac{s^{3/2} \hat{h}_2 - \hat{h}_2(1)}{1-s^2}ds + \frac{1}{8} \ln(1-u^2) \left(2 f_2(1) - \hat{g}_2'(1) \right) + \hat{h}_2(1) \ln(1+u) \\
	&+ \frac{(1+u)\ln(1+u) - u (1+2u)}{12(1+u)} \hat{\mu}^2 
	- \frac{1}{3} \hat{k} \int_0^1 \hat{b}_1^2 ds \left( \int_0^1 \hat{b}_1^2 ds \hat{k} + \hat{\mu} \right) \left(\frac{u(1+2u)}{1+u} -  \ln(1+u) \right) \\ 
	&+ \left(Li_2\left(\frac{1-u}{2}\right) + Li_2 \left( - \frac{1-u}{1+u} \right) + \frac{u}{1+u} + \frac{\ln(2)^2}{2} + \frac{1}{2} \left(\ln\left(\frac{1+u}{4} \right) -2 \right) \ln(1+u) \right) \\
	&\times  \left( \frac{1}{96} \hat{b}_1^2 (1) \left( 1 - 64 \int_0^1 \hat{b}_1^2 ds + 32 \hat{b}_1^2(1) \right) \hat{k}^2 - \frac{1}{3} \hat{b}_1^2(1) \hat{k}\hat{\mu} \right) 
\end{split}
\end{align} 
and
\begin{align}
\begin{split}
\label{E:F2}
	F_2 =& \frac{\hat{k}^2}{48} \frac{u \hat{b}_1^2 -\hat{b}_1^2(1)}{1-u^2} + \frac{2 \hat{k}^2}{3} \frac{u^2 \hat{b}_1^4 - \hat{b}_1^4(1)}{1-u^2} - \frac{2\hat{k}}{3}\left(2 \int_0^1 \hat{b}_1^2 ds \hat{k} + \hat{\mu} \right) \frac{u^2 \hat{b}_1^2 - \hat{b}_1^2(1) }{1-u^2} \\
	&+ \frac{1}{12}u^2( \hat{b}_1^{\prime})^2 + \frac{\hat{g}_2 - (u-1) \hat{g}_2'(1)}{(1-u^2)^2}
\end{split}
\end{align}
and that $x_0 = 0$.

\subsubsection{Perturbations of $g_{yz}$}
\label{A:stress-stress1}
The analysis of metric perturbations in the $yz$ direction, $h_{yz}$ is similar to that of the $xy$ components described above. In the large $Y$ limit, the homogenous version of the equation of motion for $h_{yz}$ reads
\begin{equation}
\label{E:largeYhyz}
	\partial_Y^2 h_{yz}'' + \left(\frac{\omega}{\pi T}\right)^2 h_{yz} =  \mathcal{O}(e^{-Y})
\end{equation}
with $T$ given in \eqref{E:Tdef}.
Solving \eqref{E:largeYhyz} we have
\begin{equation}
	h_{yz} = h_{yz}^{\infty} e^{\frac{i\omega}{\pi T}Y}  + \mathcal{O}(e^{-Y})\,.
\end{equation}

As before we work in the probe limit and use the notation in \eqref{E:hprobe} and \eqref{E:hexpansion}.
The equations of motion at order $\gamma^0$ read
\begin{align}
\begin{split}
	\left( (u^{-1} - u) \left(u h^{(0,n)}_{yz}\right)' \right)'  = \sigma_{yz}^{(n)}\,.
\end{split}
\end{align}
Using $\sigma_{yz}^{(0)} = \sigma_{yz}^{(1)} = 0$ and matching the small frequency solution to the large $Y$ solution we find that
\begin{equation}
	u h_{yz}^{(0,0)} = \delta(q) h_1
	\qquad
	u h_{yz}^{(0,1)} = -\frac{i}{4} h_1 \delta(q) \ln (1-u^2) 
\end{equation}
as in \eqref{E:hxyw0}.

At order $\gamma^{-1}$ the metric does not get corrected but the gauge field does get sourced by the metric. Denoting the gauge field perturbations by $\alpha_{\mu}$, i.e., $A = a dt + \omega_y b + \gamma^{-1} \alpha_{\mu} dx^{\mu}$ we obtain, following an analysis identical to the one described in the previous section, 
\begin{equation}
	\alpha_t^{(0)} = \alpha_{x}^{(0)} = 0
	\qquad
	\alpha_\pm^{(0)} = \frac{1}{2 }\hat{h}_1 \hat{b}_{1}  \delta (q \pm 2 k)\,.
\end{equation}
and
\begin{align}
\begin{split}
	\alpha_t^{(1)} &= \hat{h}_1 \left(\delta(q+2k)-\delta(q-2k)\right) \beta_7(u) \\
	\alpha_x^{(1)} & = \frac{1}{2} \hat{h}_1 \left(\delta(q-2k)+\delta(q+2k)\right)\left(\arctanh(u) \hat{b}_1^2(1) + \int_0^u \frac{\hat{b}_1^2 - \hat{b}_1^2(1)}{1-s^2} ds \right)	\\
	\alpha_{\pm}^{(1)} & = \hat{h}_1 \left( \delta(q \mp 2k) \beta_6(u) + \delta(q \pm 2k) \left(\beta_5 - \frac{1}{8} \hat{b}_1(1) \ln(1-u^2)\right) \right)
\end{split}
\end{align}
with $\beta_5$, $\beta_6$ and $\beta_7$ satisfying
\begin{subequations}
\label{E:betaieqnV3}
\begin{multline}
	\left( \left(1-u^2\right) \tilde{\beta}_5' \right)' 
	+  \hat{k} \left(\left(-\frac{1}{4u} + 8 \left(\hat{b}_1^2 - \int_0^1 \hat{b}_1^2(t) dt \right) \right) \hat{k} - 4 \hat{\mu} \right) \tilde{\beta}_5 \\
	- 4 i \hat{k} \left(\hat{b}_1 \beta_7' + 2 \beta_7 \hat{b}_1'\right)  
	= -\hat{b}_1 \ln(1-u^2) \hat{k} \left(2 \left(\hat{b}_1^2 - \int_0^1 \hat{b}_1^2(t) dt \right) \hat{k} - \hat{\mu} \right) + \frac{1}{2} u \hat{b}_1'
\end{multline}
\begin{align}
	((1-u^2)\beta_6')' +  \left( - \frac{9 \hat{k}}{4 u} +24 \hat{k}^2 \left(-\hat{b}_1^2 + \int_0^1 \hat{b}_1^2(t) dt \right) + 12 \hat{k}\hat{\mu} \right) \beta_6 + 4 i \hat{k} \left( \hat{b}_1 \beta_7' - 2 \hat{b}_1' \beta_7 \right) & = 0 \\ 
	\beta_7'' - \frac{\hat{k}}{u(1-u^2)} \beta_7 - 6 i \hat{k} \beta_6 \hat{b}_1' + 2 i \hat{k} \left( \hat{b}_1 \tilde{\beta}_5' - \tilde{\beta}_5 \hat{b}_1' \right) & = 0 
\end{align}
with
\begin{equation}
	\tilde{\beta}_5 = \beta_5-\frac{1}{8} \hat{b}_1(1) \ln(1-u^2)\,.
\end{equation}
\end{subequations}
The boundary conditions we impose on the $\beta_i$ are that they vanish at the boundary, are finite at the horizon and, in particular $\beta_7(1) = i\frac{\hat{b}_1^2(1)}{2\hat{k}}$.
Note that a possible contribution to $\alpha_{\pm}$ proportional to $\delta(q)\hat{b}_1$ vanishes due to the matching conditions with the near horizon region.

At order $\gamma^{-2}$ we obtain the set of equations
\begin{equation}
\label{E:hyzweqn}
	\left(	\left( \frac{1}{u} - u \right)  \left( u h_{yz}^{(2,n)} \right)^\prime \right)' - \frac{4 q^2}{u} h_{yz}^{(2,n)} = \sigma_{yz}^{(n)} \,.
\end{equation}
The solution to \eqref{E:hyzweqn} for $n=0$ and $n=1$ is given by 
\begin{equation}
	h_{yz}^{(2,0)} = 0
\end{equation}
and
\begin{align}
\begin{split}
 	h_{yz}^{(2,1)} = & \frac{i \hat{h}_1}{u}   \delta(q) \left(F_3 + \int_0^s \frac{s \int_s^1 F_4(s) dt}{1-s^2} ds \right)  \\
	&-  \frac{i \hat{h}_1}{8 u}   \left(\delta(q-4k) + \delta(q+4k)\right) \left(\mathfrak{c}_1 \int_u^1 F_5 \mathfrak{c}_2 ds + \mathfrak{c}_2 \int_0^u F_5 \mathfrak{c}_1 ds \right)  
\end{split}
\end{align}
where
{\small
\begin{align}
\begin{split}
\label{E:F3}
	F_3 =& \ln \left(1-u^2\right) \left(-\frac{1}{6} \hat{k} \hat{\mu } \left(\int_0^1 \hat{b}_1^2(t) \, dt + \hat{b}_1^2(1)\right) + \frac{1}{8} \left(2 f_2(1)-\hat{g}_2'(1)\right)-\frac{\hat{\mu}^2}{24} \right) \\
	&+\ln \left(1-u^2\right) \left(\frac{1}{192} \hat{k}^2 \left(-64 \hat{b}_1^2(1)  \int_0^1 \hat{b}_1^2(t) \, dt -32 \left(\int_0^1 \hat{b}_1^2(t) \, dt\right)^2+\hat{b}_1^2(1) \left(32 \hat{b}_1^2(1)-1+\log (4)\right)\right)\right) \\
	%
   %
   &-\text{Li}_2\left(\frac{u-1}{u+1}\right) \left(\frac{1}{3} \hat{k} \hat{\mu } \left(\hat{b}_1^2(1)-\int_0^1 \hat{b}_1^2(t) \, dt\right)-\frac{1}{3} \hat{k}^2 \left(\int_0^1 \hat{b}_1^2(t) \, dt-\hat{b}_1^2(1)\right){}^2+\frac{\hat{g}_2' (1)}{4}\right) \\
   &-
   \frac{\hat{k}^2}{384} \hat{b}_1^2(1)
   \left(256 u \left(\hat{b}_1^2(1)-2 \int_0^1
   \hat{b}_1^2(t) \, dt\right)+\ln
   ^2\left(1-u^2\right)\right) \\
   &
   -\frac{\hat{k}^2}{3} \tanh
   ^{-1}(u) \left(\ln \left(1-u^2\right)
   \left(\int_0^1 \hat{b}_1^2(t) \,
   dt\right){}^2+2 \ln \left(\frac{2}{u+1}\right)
   \left(\int_0^1 \hat{b}_1^2(t) \,
   dt-\hat{b}_1^2(1)\right)^2\right) \\
   &-\frac{\hat{k}^2}{3} \tanh^{-1}(u)
   \left(4 \hat{b}_1^2(1)
   \left(\int_0^1 \hat{b}_1^2(t) \, dt\right)-2
   \hat{b}_1(1)^4\right) \\
   &-\frac{\hat{k}^2}{6} \left( \left(\ln
   ^2(1-u)-\ln ^2(u+1)\right) \left(\int_0^1
   \hat{b}_1^2(t) \, dt\right)^2
   -\frac{1}{3}
   \tanh ^{-1}(u)^2 \left(\int_0^1 \hat{b}_1^2(t)
   \,
   dt-\hat{b}_1^2(1)\right)^2\right) \\
   &-\frac{1}{3} \hat{k} \hat{\mu } \left(\tanh
   ^{-1}(u) \left(\left(-2 \ln (u+1)+\tanh
   ^{-1}(u)+\ln (4)\right) \left(\int_0^1
   \hat{b}_1^2(t) \, dt-\hat{b}_1^2(1)\right)+2
   \hat{b}_1^2(1)\right)-2 \hat{b}_1^2(1)
   u\right) \\
   &-\frac{1}{4} \hat{g}_2'(1)
   \left(\frac{1}{u+1}-\tanh ^{-1}(u) \left(-2
   \ln (u+1)+\tanh ^{-1}(u)-1+\ln
   (4)\right)\right)\\
   &+\frac{1}{48} \hat{\mu }^2
   \left(-4
   \text{Li}_2\left(\frac{1-u}{2}\right)+\ln
   ^2(u+1)+\ln (1-u) \left(\ln
   \left(-\frac{16}{u-1}\right)-2 \ln
   (u+1)\right)\right)
\end{split}
\end{align}
}
and
\begin{align}
\begin{split}
\label{E:F4}
	F_4 =& \ln \left(1-u^2\right) \left(-\frac{1}{8} \left(u^2-1\right) \hat{b}_1'(u)^2+\frac{\hat{k}^2 \hat{b}_1(u)^2}{32 u}-\frac{\hat{b}_1(1) \hat{k}^2 \hat{b}_1(u)}{32 u}\right)-\left(u^2-1\right) \hat{b}_1'(u) \beta_5'(u) \\
	&+\frac{u \left(u^2-1\right)
   \hat{b}_1'(u) \left(u \hat{b}_1'(u)+3
   \hat{b}_1(1)\right)-2 \hat{\mu }^2
   \left(u^2-1\right)}{12
   \left(u^2-1\right)} \\
   &+\left(\frac{4 \hat{k}^2 u^2
   \left(\hat{b}_1(u)^2-\hat{b}_1(1)^2\right)}{3
   \left(u^2-1\right)}+\frac{2 \hat{k} \hat{\mu }
   \left(u^2-1\right)}{3-3 u^2}\right)
   \int_0^1 \hat{b}_1(t)^2 \,
   dt \\
   &+\frac{2 \hat{k}^2 \left(u^2-1\right)
   \left(\int_0^1 \hat{b}_1(t)^2 \,
   dt\right){}^2}{3-3 u^2}+\frac{\hat{k} u
   \left(\hat{b}_1(u)^2-\hat{b}_1(1)^2\right)
   \left(32 \hat{\mu } u-\hat{k}\right)}{48
   \left(u^2-1\right)} \\
   &+\frac{2 \hat{k}^2 u^2
   \left(\hat{b}_1(u)^4-\hat{b}_1(1)^4\right)}{3(1- u^2)}+\frac{\hat{k}^2 \hat{b}_1(u)
   \beta_5(u)}{4 u}+\frac{-(u-1)
   \hat{g}_2'(1)+\hat{g}_2(u)-\hat{g}_2(1)}{\left(u^2-1\right)^2} 
\end{split}
\end{align}
and $\mathfrak{c}_i$ and $F_5$ are given in \eqref{E:cequation} and \eqref{E:Qval}

\section{Numerical evaluation of the remaining correlators}
\label{A:Numerical}
In section \ref{S:Perturbative} we have evaluated the coefficients $s_i$, $u_i$ and $t_i$ defined in \eqref{E:si}, \eqref{E:ui1}, \eqref{E:ui2} and \eqref{E:ti} numerically omitting some details of the computation. In what follows we will provide an extended description of this computation. 

In order to compute $s_2$, $u_2$, $u_5$ and $t_2$ we need a handle on the functions $\beta_i$ defined in \eqref{E:betaieqn}, \eqref{E:betaieqnV2} and \eqref{E:betaieqnV3}. To compute the $\beta_i$'s in the small $\Delta$ limit elaborated on in sections \ref{S:PerturbativeBgd} and \ref{S:Perturbative} we define
\begin{equation}
	\beta_i = \sum_{n=0} \beta_i^{(n)} \Delta^n
\end{equation}
and solve the equations of motion for $\beta_i^{(n)}$ order by order in $\Delta$. Fortuntaely, all the equations of motion decouple in the small $\Delta$ limit and can be solved using Green functions.

Let us denote by $\mathfrak{a}_i(u;\kappa)$ the solutions to 
\begin{equation}
	\mathfrak{a}_i'' - \frac{\mathfrak{a}_i \kappa^2}{u (1-u^2)} = 0
\end{equation}
where $\mathfrak{a}_1(0) = 0$, $\mathfrak{a}_2(0)=1$, and $\mathfrak{a}_2(1)=0$ while $\mathfrak{a}_1$ is finite and non zero at the horizon. Similarly, let us denote by $\mathfrak{p}_i(u;\kappa)$ the solutions to 
\begin{equation}
	\left((1-u^2) \mathfrak{p}_i'\right)' + \left(8 \kappa \hat{\mu}_0 - \frac{\kappa^2}{u} \right) \mathfrak{p}_i = 0
\end{equation}
with $\mathfrak{p}_1(0)=0$ and $\mathfrak{p}_2(0)=1$, $\mathfrak{p}_2$ finite at $u=1$ while $\mathfrak{p}_1$ diverges logarithmically at the horizon.
The Green function for the $\beta_i$ are given by $\mathfrak{p}_i$ and $\mathfrak{a}_i$. We find
\begin{align}
\begin{split}
	\beta_1 =&  \Bigg(i \hat{k}_0 \mathfrak{a}_1(u;\hat{k}_0/2) \int_u^1 \frac{ \hat{b}_{1,1}(s) \mathfrak{a}_2(s;\hat{k}_0/2)}{1-s^2} ds + i \hat{k}_0 \mathfrak{a}_2(u;\hat{k}_0/2) \int_0^u \frac{\hat{b}_{1,1}(s) \mathfrak{a}_1(s;\hat{k}_0/2)}{1-s^2} ds \\
		&+ \frac{4 i \hat{b}_1(1)}{\hat{k}_0} \frac{\mathfrak{a}_1(u;\hat{k}_0/2)}{\mathfrak{a}_1(1;\hat{k}_0/2)} \Bigg) \Delta + \mathcal{O}(\Delta^2) \\
	\beta_2 =& 4 i \hat{k}_0 \Bigg(\mathfrak{p}_1(u;\hat{k}_0) \int_u^1 \mathfrak{p}_2(s;\hat{k}_0) \left(\hat{b}_{1,1}(s) \beta_1'(s) -\hat{b}_{1,1}'(s) \beta_1(s)\right) ds \\
		&+ \mathfrak{p}_2(u;\hat{k}_0) \int_0^u \mathfrak{p}_1(s;\hat{k}_0)  \left(\hat{b}_{1,1}(s) \beta_1'(s) - \hat{b}_{1,1}'(s) \beta_1(s)\right) ds \Bigg) \Delta + \mathcal{O}(\Delta^3) \\
	\beta_3 =&   \Bigg(-\frac{i \hat{k}_0}{8}\mathfrak{a}_1(u;\hat{k}_0/2)  \int_u^1 \frac{ \hat{b}_{1,1}(s) \mathfrak{a}_2(s;\hat{k}_0/2) }{s(1-s^2)} ds -\frac{i \hat{k}_0}{8} \mathfrak{a}_2(u;\hat{k}_0/2)  \int_0^u \frac{\hat{b}_{1,1}(s) \mathfrak{a}_1(s;\hat{k}_0/2) }{s(1-s^2)} ds \\
	&- \frac{i \hat{b}_{1,1}(1)}{2 \hat{k}_0} \frac{\mathfrak{a}_1(u;\hat{k}_0/2) }{\mathfrak{a}_1(1;\hat{k}_0/2) } \Bigg) \Delta + \mathcal{O}(\Delta^2) \\
	\beta_4 =& \Bigg(\mathfrak{p}_1(u;\hat{k}_0) \int_u^1 \mathfrak{p}_2(s;\hat{k}_0) \left( 4 i \hat{k}_0\left(\hat{b}_{1,1}(s) \beta_1'(s) -\hat{b}_{1,1}'(s) \beta_1(s)\right) -(\hat{b}_{1,1}^2(s))' \right)ds \\
		&+ \mathfrak{p}_2(u;\hat{k}_0) \int_0^u \mathfrak{p}_1(s;\hat{k}_0)   \left( 4 i \hat{k}_0\left(\hat{b}_{1,1}(s) \beta_1'(s) -\hat{b}_{1,1}'(s) \beta_1(s)\right) -(\hat{b}_{1,1}^2(s))' \right)\Bigg) \Delta + \mathcal{O}(\Delta^3)  \\
	\beta_5 =& \left(\frac{1}{8} \left(\hat{b}_{1,1}(1)-\hat{b}_{1,1}(u)\right) \ln(1-u^2) +  \frac{\hat{b}_{1,1}(u)}{32\hat{k}_0 \hat{\mu}_0} \right) \Delta + \mathcal{O}(\Delta^3) \\ 
	\beta_6 = & -4 i \hat{k}_0 \Bigg(\mathfrak{p}_2\left(u;3\hat{k}_0/2\right) \int_0^u \mathfrak{p}_1\left(s;3\hat{k}_0/2\right) (2 \beta_{7}(s) \hat{b}_{1,1}'(s) - \beta_7'(s) \hat{b}_{1,1}(s)) ds \\
		&+ \mathfrak{p}_1\left(u;3\hat{k}_0/2\right) \int_u^1  \mathfrak{p}_2\left(s;3\hat{k}_0/2\right) (2 \beta_{7}(s) \hat{b}_{1,1}'(s) - \beta_7'(s) \hat{b}_{1,1}(s)) ds \Bigg) \Delta + \mathcal{O}(\Delta^4) \\
	\beta_7  =& \Bigg( \frac{1}{2} i \hat{k}_0 \left( \mathfrak{a}_2(u;\hat{k}_0) \int_0^u \frac{s \hat{b}_{1,1}^2(s) \mathfrak{a}_1(s;\hat{k}_0)}{1-s^2}ds + \mathfrak{a}_1(u;\hat{k}_0) \int_u^1 \frac{s \hat{b}_{1,1}^2(s) \mathfrak{a}_2(s;\hat{k}_0)}{1-s^2} ds \right) \\
		& + \frac{i \hat{b}_{1,1}^2(1) \mathfrak{a}_1(u;\hat{k}_0)}{2 \hat{k}_0 \mathfrak{a}_1(1;\hat{k}_0)}  \Bigg)\Delta^2 + \mathcal{O}(\Delta^4)\,.
\end{split}
\end{align}
Note that $\beta_6$ is of order $\Delta^3$.
By evaluating $\mathfrak{a}_i$ and $\mathfrak{p}_i$ numerically we obtain \eqref{E:sunumerical} and \eqref{E:t2numerical}.

\section{Heun polynomials}
\label{A:analyticzeromode}
In section \ref{S:Instability} we indicated that for special values of $\hat{k}$ and $\hat{\mu}$,  \eqref{E:eigenequation} possess polynomial solutions called Heun polynomials. To find these solutions we start by solving \eqref{E:eigenequation} using a series expansion near $u=0$
\begin{equation}
	\beta = u\sum_{n=0}^{\infty}d_{n}u^{n}
	\label{E:powerseries}
\end{equation}
with $d_0=1$.
Inserting the series expansion into \eqref{E:eigenequation} yields the following recursion relation for $d_n$
\begin{equation}
	4d_{n-1}\left(-4\hat{k}\hat{\mu}+n^{2}+n\right)+\hat{k}^{2}d_{n}-4\left(n^{2}+3n+2\right)d_{n+1} = 0 .
	\label{E:recursionRelation}
\end{equation}
The general solution to \eqref{E:recursionRelation} is given by
\begin{equation}
d_{n}= d_{0}\frac{1}{2^{2n}n!\left(n+1\right)!}\left(\hat{k}^{4q(n)} + \sum_{i=0}^{q(n) - 1}C_{q(n)}^{i}\hat{k}^{4i}\right)
\label{E:recursionSolution}
\end{equation}
where $n > 1$ and
\begin{align}
q(n)=\begin{cases}
\frac{n}{2} & \mbox{n even}\\
\frac{n-1}{2} & \mbox{n odd}
\end{cases}
\end{align}
and
\begin{align}
\begin{split}
C_{q(n)}^{i}	&=	2^{4\left(q(n)-i\right)}\sum_{m_{1}=0}^{\tilde{d}(i,n)}\sum_{m_{2}=m_{1}}^{\tilde{d}(i,n)}...\\
		&...\sum_{m_{\tilde{d}}=m_{\tilde{d}-1}}^{\tilde{d}(i,n)}\prod_{p=1}^{q(n)-i}\left(m_{p}+\left(2p-1\right)\right)\left(m_{p}+2p\right)\left[\left(m_{p}+\left(2p-1\right)\right)\left(m_{p}+2p\right)-4\hat{k}\hat{\mu}\right]
\end{split}
\end{align}
with
\begin{equation}
	\tilde{d}(i,n) = \begin{cases}
		2i & \mbox{n even}\\
		2i+1 & \mbox{n odd}\,,
\end{cases}
\end{equation}
as can be verified by direct insertion into \eqref{E:recursionRelation} \cite{Omrie:Thesis}.

A polynomial solution of order $l+1$ corresponds to a solution of \eqref{E:recursionRelation} with $d_m = 0$ for all $m>l$. Such a solution exists if and only if
\begin{align}
\begin{split}
	d_{l} \left(-4 \hat{k} \hat{\mu }+\left(l+1\right)^2+l+1\right)=0\\
	\hat{k}^2 d_{l}+4 d_{l-1} \left(-4 \hat{k} \hat{\mu
   }+l^2+l\right) = 0 
\end{split}
\end{align}
or, equivalently,
\begin{align}
\begin{split}
\label{E:critkmu}
	\hat{k} \hat{\mu } &= \frac{l^2+3 l+2}{4}\\
	\hat{\mu }^2 &= \frac{\left(l+1\right) \left(l+2\right)^2 }{128 }\frac{d_{l}}{d_{l-1}} . 
\end{split}
\end{align}
Thus, to obtain a polynomial solution to \eqref{E:eigenequation} we solve \eqref{E:critkmu} and \eqref{E:recursionRelation}. The first ten Heun polynomials obtained using this method are given in table  \ref{table:Heun polynomials} and equations \eqref{E:HeunSols}. 
\begin{table}[ht] 
\centering 
\begin{tabular}{l l l} 
\hline\hline
Heun polynomial & $\hat{k}$ & $\hat{k}\hat{\mu}$ \\ [1ex] 
\hline
$u(1+\sqrt{2}u)$ & $2^{7/4}$ & $\frac{3}{2}$\\ 
$u\left(1+\sqrt{14} u+3 u^2\right)$ & $2\times2^{3/4}\sqrt[4]{7}$ & $3$ \\ 
$u(1+7.04624 u+13.5498 u^2+7.69195 u^3)$ & 7.50799 & 5 \\ 
$P_4(u)$ & 9.52666 & 7.5 \\
$P_5(u)$ & 11.538 & 10.5 \\
$P_6(u)$ & 13.5457 & 14 \\
$P_7(u)$ & 15.5513 & 18 \\
$P_8(u)$ & 17.5555 & 22.5 \\
$P_9(u)$ & 19.5588 & 27.5 \\
$P_{10}(u)$ & 21.5614 & 33 \\ [1ex] 
\hline 
\end{tabular} 
\caption{\label{table:Heun polynomials} The ten lowest order Heun polynomials solutions for \eqref{E:eigenequation} obtained by solving \eqref{E:critkmu} and \eqref{E:recursionRelation} and their appropriate values of $\hat{k}$ and $\hat{\mu}$. The explicit form of some of the longer polynomials appear in equation \eqref{E:HeunSols}.}
\end{table} 
\begin{align}
\label{E:HeunSols}
\begin{split}
	P_4(u) =& u (1 + 11.3446 u + 38.2337 u^2 + 49.602 u^3 + 21.8614 u^4) \\
	P_5(u) =& u (1 + 16.6408 u + 85.6386 u^2 + 187.593 u^3 + 183.712 u^4 + 66.2392 u^5) \\
	P_6(u) =& u (1 + 22.9358 u + 166.351 u^2 + 540.333 u^3 + 873.328 u^4 + 686.967 u^5  \\ & + 209.662 u^6) \\
	P_7(u) =& u (1 + 30.2303 u + 292.956 u^2 + 1309.76 u^3 + 3080.57 u^4 + 3938.17 u^5 \\&+ 2588.58 u^6 + 685.03 u^7) \\
	P_8(u) =& u (1 + 38.5243 u + 480.041 u^2 + 2812.54 u^3 + 8962.95 u^4 + 16456.8 u^5 \\&+ 17385.7 u^6 + 9814.6 u^7 + 2292.87 u^8) \\
	P_9(u) =& u (1 + 47.8181 u + 744.19 u^2 + 5516.54 u^3 + 22732.5 u^4 + 55918.8 u^5 \\&+ 84030 u^6 + 75604.1 u^7 + 37401 u^8 + 7821.52 u^9) \\
	P_{10}(u) =& u (1 + 58.1117 u + 1103.99 u^2 + 10082.3 u^3 + 51966 u^4 + 163682 u^5 \\&+ 326741 u^6 + 415064 u^7 + 325110 u^8 + 143128 u^9 + 27092.7 u^{10})
\end{split}
\end{align}

A generic point on the critical curve displayed in the left panel of figure \eqref{F:critcurve} is described by a Heun function which may be represented by the infinite power series  \eqref{E:powerseries} which converges at $u = 1$. Heun polynomials are special points where this power series terminated at some finite order. The first few stable Heun polynomial solutions can also be seen in figure \ref{F:critcurve}. 

The careful reader will note that for $l>3$ there is more than one real $\hat{k}$ and $\hat{\mu}$ which will solve the polynomial equation \eqref{E:critkmu}. 
For example,  
\begin{align}
\begin{split}
\beta(u) = u(1 + 2.08578 u -1.54983 u^2 -2.97219 u^3 )\\
\end{split}
\label{E:non stable poly example}
\end{align}
with $\hat{k} = 4.08488,\, \hat{k}\hat{\mu} = 5$ solves \eqref{E:eigenequation}.
In table \ref{table:Heun polynomials} we have listed, for each value of $l$, the solution to \eqref{E:critkmu} with the highest value of $\hat{k}^2$ corresponding to a Heun polynomial with no roots on the interval $0<u<1$. Thus, the solution \eqref{E:non stable poly example} has been removed since it vanishes at  $u=0.824$.

The fact that there exist values of $l$ for which \eqref{E:eigenequation} has more than one solution is not surprising. Recall that for every value of $\hat{k}\hat{\mu}$, equation \eqref{E:eigenequation} toghether with the boundary conditions, $ \beta(0) = 0, \beta(1) = \mbox{finite}$, may be considered as a Sturm---Liouville system with eigenvalue $\lambda \equiv -\hat{k}^2$. 
From Sturm---Liouville theory we are guaranteed that for each $\hat{k}\hat{\mu}$ there is an infinite set of real eigenvalues $\lambda_{n}$ such that $\lambda_{n}\rightarrow\infty$ as $n\rightarrow\infty$. Moreover, if we order the eigenvalues from smallest to largest, $\lambda_{0}<\lambda_{1}<...$ then each $\lambda_{n}$ corresponds to a different eigenfunction that has exactly $n$ zeros in the interval $0<u<1$. In this language table \ref{table:Heun polynomials} lists only the smaller, $\lambda_0$, eigenvalues and eigenfunctions. All the solutions for which $\lambda_i < 0$ with $i \geq 1$ will necessarily lie inside the critical curve described in the left panel of figure \ref{F:critcurve} and are expected to be unstable due to their higher free energy (see e.g., \cite{Winstanley:2008ac}).

\end{appendix}	
\bibliographystyle{JHEP}
\bibliography{Modulatedbib}

\end{document}